%% file: _main.tex
\input{_constants}
\arxiv 

\pdfoutput=1
\documentclass[10pt,twocolumn,letterpaper]{article}
\input{cvpr_header}

\begin{document}
\title{\paperTitle}
\author{\authorBlock}


\twocolumn[{%
\maketitle
\begin{center}
    \centering
    \captionsetup{type=figure}
     \includegraphics[width=\linewidth]{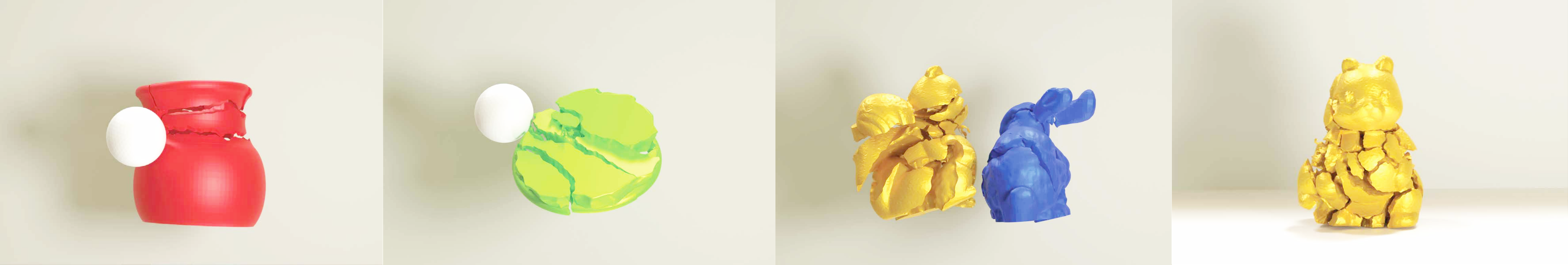}
      \vspace{-5mm}
      \caption{
      We introduce a novel learning-based approach for generating brittle fracture animations integrated with rigid-body simulations. This approach trains generative models for each shape using brittle fracture simulation data. Using our method, we can predict brittle fracture patterns rapidly with impulse information. This enables the seamless continuation of rigid-body simulations when detecting collision detection.
    }
\end{center}%
}]



\input{00_abstract}
\input{01_intro}

\input{02_related}

\input{03_method}

\input{10_conclusion}

{\small
\bibliographystyle{ieee_fullname}
\bibliography{11_references}
}


\end{document}


\title{\paperTitle}
\author{\authorBlock}
\maketitlesupplementary

\input{12_appendix}

{\small
\bibliographystyle{ieee_fullname}
\bibliography{11_references}
}

%% file: _constants.tex
\def\paperID{0000}
\def\confName{CVPR}
\def\confYear{2024}

\def\paperTitle{DeepFracture: A Generative Approach for Predicting Brittle Fractures with Neural Discrete Representation Learning}

\def\authorBlock{
    Yuhang Huang$^\dagger$ \qquad
    Takashi Kanai$^\ddagger$ \\
    The University of Tokyo \\
    {\tt\small $\dagger${}nikoloside@gmail.com , $\ddagger${}kanait@acm.org }
}

\newif\ifreview 
\newif\ifarxiv \newcommand{\arxiv}{\arxivtrue}
\newif\ifcamera 
\newif\ifrebuttal 

%% file: cvpr_header.tex
\ifreview \usepackage[review]{cvpr} \fi
\ifarxiv \usepackage[pagenumbers]{cvpr} \fi
\ifrebuttal \usepackage[rebuttal]{cvpr} \fi
\ifcamera \usepackage{cvpr} \fi

\input{_macros}  

\usepackage{xr-hyper}

\makeatletter
\newcommand*{\addFileDependency}[1]{
  \typeout{(#1)}
  \@addtofilelist{#1}
  \IfFileExists{#1}{}{\typeout{No file #1.}}
}

\makeatother

\definecolor{cvprblue}{rgb}{0.21,0.49,0.74}
\usepackage[pagebackref,breaklinks,colorlinks,citecolor=cvprblue]{hyperref}
\usepackage[capitalize]{cleveref}
\crefname{section}{Sec.}{Secs.}
\crefname{table}{Table}{Tables}
\crefname{figure}{Fig.}{Figs.}

%% file: _macros.tex
\usepackage{graphicx}	
\usepackage{amsmath}	
\usepackage{amssymb}	
\usepackage{booktabs}
\usepackage{times}
\usepackage{microtype}
\usepackage{epsfig}
\usepackage[table,xcdraw,dvipsnames]{xcolor}
\usepackage{caption}
\usepackage{float}
\usepackage{placeins}
\usepackage{color, colortbl}
\usepackage{stfloats}
\usepackage{enumitem}
\usepackage{tabularx}
\usepackage{xstring}
\usepackage{multirow}
\usepackage{xspace}
\usepackage{url}
\usepackage{subcaption}
\usepackage{xcolor}
\usepackage[hang,flushmargin]{footmisc}

\usepackage{amsfonts} 
\usepackage{bbm}
\usepackage{algorithm}
\usepackage{algpseudocode}

\ifcamera \usepackage[accsupp]{axessibility} \fi

% ...

\ifarxiv  \fi

\newcommand{\R}[1]{{%
    \textbf{%
        \ifstrequal{#1}{1}{\textcolor{red}{R#1}}{%
        \ifstrequal{#1}{2}{\textcolor{blue}{R#1}}{%
        \ifstrequal{#1}{3}{\textcolor{magenta}{R#1}}{%
        \ifstrequal{#1}{4}{\textcolor{teal}{R#1}}{%
                           \textcolor{cyan}{R#1}%
        }}}}%
    }%
}}

\def\mathbi#1{\textbf{\em #1}}

%% file: 00_abstract.tex
\begin{abstract}

In the field of brittle fracture animation, generating realistic destruction animations using physics-based simulation methods is computationally expensive. While techniques based on Voronoi diagrams or pre-fractured patterns are effective for real-time applications, they fail to incorporate collision conditions when determining fractured shapes during runtime.

This paper introduces a novel learning-based approach for predicting fractured shapes based on collision dynamics at runtime. Our approach seamlessly integrates realistic brittle fracture animations with rigid body simulations, utilising boundary element method (BEM) brittle fracture simulations to generate training data. To integrate collision scenarios and fractured shapes into a deep learning framework, we introduce generative geometric segmentation, distinct from both instance and semantic segmentation, to represent 3D fragment shapes.

We propose an eight-dimensional latent code to address the challenge of optimising multiple discrete fracture pattern targets that share similar continuous collision latent codes. This code will follow a discrete normal distribution corresponding to a specific fracture pattern within our latent impulse representation design.
This adaptation enables the prediction of fractured shapes using neural discrete representation learning. Our experimental results show that our approach generates considerably more detailed brittle fractures than existing techniques, while the computational time is typically reduced compared to traditional simulation methods at comparable resolutions.

\end{abstract}

%% file: 01_intro.tex
\section{Introduction}
\label{sec:introduction}

Brittle fracture animations provide impressive visual effects to video games, movies, and virtual reality. 
Most simulation methods do not focus on crack propagation but instead on determining the cutting meshes.
Recently, physics-based simulation methods based on quasi-static crack propagation generate detailed realistic fractured shapes and surfaces. 
However, current physics-based simulation methods based on crack propagation \cite{hahn2016fast, brittlempm2022}, including those used in the film industry, are computationally expensive.

In real-time applications like virtual reality or games, a more popular alternative involves creating a pre-fractured pattern during the modelling stage and swapping from the original shape to the fractured shape upon collision. 
Several methods \cite{muller2013real, schvartzman2014fracture} dynamically determine the number and size of destruction fragments based on the Voronoi diagram in advance. 
However, the monotonous Voronoi-like shapes make it difficult to represent complex real-world fracture patterns.

To address the challenge of recognising pre-fractured patterns in materials with weak structures, Sell\'{a}n et al.\cite{breakinggood2023} introduced a technique for generating pre-fractured shapes blended from multiple worst-case structural analyses and employing pre-fractured shapes at run-time.
However, while it significantly improves the quality of shapes with weak structures, it has limitations in representing crack propagation shapes inside the object.

In this paper, we introduce a novel approach for predicting brittle fracture patterns using neural discrete representation learning.
We reconceptualise the challenge of brittle fracturing as predicting a specific fracture pattern related to a BEM simulation collision condition, treating it as a conditional fracture pattern 3D shape prediction.
Our method consists of two main processes: 
1) During the learning process, our framework uses the BEM brittle fracture simulation \cite{hahn2016fast} to generate training data reflecting collision scenarios and their resultant fracture patterns for training the generative model.
2) During the run-time process, we predict the fractured surfaces with the generative model, then synthesize the fractured surfaces and original meshes and use the reconstructed shapes in the physics engine. 
Our technical contributions are:
\begin{itemize}
\item We introduce the 3D fracture representation of generative geometric segmentation and discuss the appropriate learning format (GS-SDF) and suitable networks.

\item We design a latent impulse representation and decoder-only neural discrete representation learning framework to optimise multiple discrete fracture patterns with a continuous latent impulse code.

\item For the run-time reconstruction, we propose a novel SDF-based cage-cutting method to preserve the external mesh of the original shape in the generated fragments.

\item We first successfully constructed a deep learning framework for fracture animation that generates a specific fracture pattern tailored to each collision with a variable number of segments.

\end{itemize}

%% file: 02_related.tex
\section{Related Work}
\label{sec:related}

\subsection{Physics-based simulation}

Terzopoulos and Fleischer were pioneers in proposing a fracture model in the realm of CG \cite{terzopoulos1988modeling}. 
Subsequently, various methods emerged, including those based on spring-mass systems \cite{norton1991animation,mazarak1999animating}, FEM \cite{o1999graphical,o2002graphical,bao2007fracturing}, XDEM \cite{imagire2009a, imagire2009b}, and graph-based FEM \cite{mandal2023remeshing}.
These physics-based simulation methods do not focus on crack propagation but rather on determining the cutting meshes.

Hahn and Wojtan \cite{hahn2015high,hahn2016fast} and Zhu et al. \cite{zhu2015simulating} initially introduced quasi-static crack propagation-based BEM brittle fracture simulation into CG. Limited to cracks starting on surface meshes, Zhu et al. \cite{zhu2015simulating} proposed a faster BEM-based crack propagation algorithm without volumetric sampling. Following the main concept of crack propagation simulation, methods using XFEM \cite{chitalu2020}, and MPM based on the phase-field approach \cite{brittlempm2022} have been developed.
Notably, the fracture simulation using BEM \cite{hahn2015high} provides a stable fracturing method that can be used in various breaking scenarios, which employs mesh-only in the BEM solver and a voxel grid in generating a new surface. The subsequent work \cite{hahn2016fast} improves the simulation speed and integrates their framework with impulse-based rigid-body simulation. 
This allows for more efficient and flexible fracture surface generation. However, crack propagation-based simulations still require enormous computational time.

\subsection{Real-time Aware Animation}

Real-time-aware methods typically focus on two key aspects individually: 1) creating pre-fracture shapes geometrically for physics dynamics rigid-body system, and 2) describing the relationship between external force loads applied to the fractured object and the resulting fractured pieces.

Among geometrically-based methods for fracture animation, Voronoi diagram-based techniques \cite{neff1999visual, AURENHAMMER2000201, raghavachary2002fracture} serve as a rapid alternative for determining fractured shapes instead of relying on physics-based approaches.
In contrast, various methods predefine both the fracture pattern and the number of fractured pieces \cite{su2009energy, hellrung2009geometric, liu2011meshless, muller2013real, schvartzman2014fracture}.
Sell\'{a}n et al. recently introduced a technique generating weak structure-aware pre-fractured shapes through worst-case structural analysis \cite{breakinggood2023}. 
However, these pre-fracture patterns lack brittle fracture propagation. They work well in thin shapes but are not inadequate in interior fractured surfaces.

\subsection{Fracture Animation with Data-Driven Approach}

The data-driven approach utilises a pre-processing process to save computational cost at run-time. 
A data-driven method \cite{schvartzman2014fracture} was introduced that utilises RBF networks to optimise a 3D Centroid Voronoi Diagram-based segmentation with external forces.
\cite{kanailab2018} utilises the random regression forest method to predict the crack-propagation surfaces based on the fracture framework of \cite{hahn2015high}.
However, this method cannot reduce the main time-consuming part of mesh updates.

\begin{figure*}[tb]
  \centering
 \mbox{} \hfill
  \includegraphics[width=\linewidth]{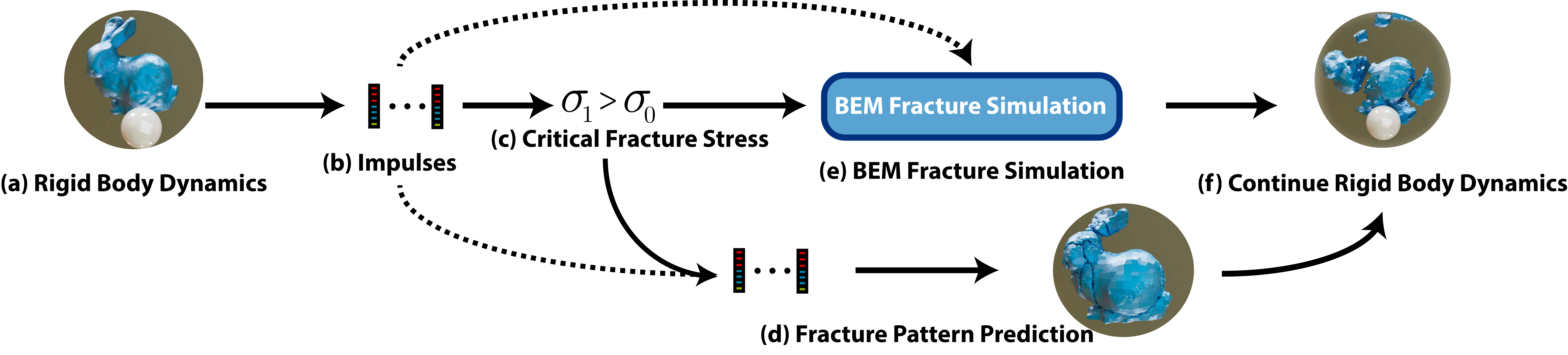}
  \vspace{-5mm}
 \hfill \mbox{}
  \caption{Overview of our framework: The data generation pipeline is composed of (a) Rigid-Body Dynamics, (b) Impulses during Collision, (c) Solving Fracture Critical Stress for judging whether break or not, (e) BEM Fracture Simulation with Impulse-based Rigid Body Engine \cite{hahn2016fast}, (f) Continue Rigid-Body Dynamics; Our work is focus on using (d) Fracture Pattern Prediction to replace (e) process.
  \label{fig:overview-new} }
  \vspace{-5mm}
\end{figure*}

\subsection{3D Representations and Related Fracture Tasks in Deep Learning}

The initial 3D generative networks were grounded in voxel-based representation \cite{wu2016learning}.
Among various generative network techniques, neural discrete representation learning \cite{van2017neural} introduces a novel generative approach to encode the latent vector discretely.
Implicit function-based 3D representation techniques \cite{park2019deepsdf} have been refined for two-piece fracture repair \cite{lamb2022mendnet, lamb2022deepjoin, lamb2023fantastic}. 
Multi-part or fractured shapes assembly is another work on fractured shapes in deep learning. Sell\'{a}n et al. \cite{NEURIPS2022_fe18f209} presents a database of geometric segmentation for assembly tasks. 
Geometric segmentation, a novel concept introduced in \cite{NEURIPS2022_fe18f209}, differs from other instance segmentation tasks because it does not assume semantic information. 

Generating fracture patterns with a variable number of geometric segments is related to, but different from, these two tasks. Therefore, we must develop our own dataset, adapted to our task of predicting conditional generative geometric segmentation with the conditional collision scenario generated by simulation.

%% file: 03_method.tex
\section{Overview}
\label{sec:overview}

Our method simulates scenarios where objects undergo collisions using impulse-based rigid-body simulation engines like Bullet Physics \cite{coumans2015bullet}. As shown in Figure~\ref{fig:overview-new}, our method and the method of Hahn and Wojtan \cite{hahn2016fast} share the same rigid-body physics dynamics and address crack propagation in the rigid-body system, treating it as a momentary process.
Instead of relying on computationally intensive brittle fracture simulations focused on crack propagation, we utilise a deep learning approach to predict fracture patterns.
As shown in Figure~\ref{fig:overview-new}, when a collision occurs during runtime, the predicted fracture pattern is immediately applied to the impacted object by swapping the result of the process (e) in Figure~\ref{fig:overview-new} with the process (d) in Figure~\ref{fig:overview-new}. 
We abstract fracturing scenarios caused by collisions to predict the fracture patterns and encode the scene into a deep learning-friendly dataset. This dataset pairs impulse data inputs with outputs of fractured fragments. We have designed a custom conditional generative model to enable the generation of these fracture patterns at various 3D resolutions through a discrete representation training framework.

In this paper, we first discuss brittle fracture simulation and deep learning modules in the learning process in Section \ref{sec:learning-process}. We then outline our runtime procedures in Section \ref{sec:run-time-process}, followed by a detailed presentation of our experiments and discussions in Section \ref{sec:experiment}.

\subsection{Impulse Based Brittle Fracture Physics}
\label{sec:prerequisite}

We use the impulse-based rigid body simulation to build our framework.
As shown in Figure~\ref{fig:overview-new} (a)-(b), when a collision occurs, there are discrete impulses $J$ created by the same rigid-body dynamics process both from our method and simulation method \cite{hahn2016fast}.
The formula can be illustrated as:
\begin{equation}
\label{equ:impulse}
\begin{split}
\forall &\mathbi{V}^{\mathrm{raw}}_i = [\mathbi{p}_i\: {\mathbi{d}_i}\: I_i], \quad  \exists(A_i,\mathbi{n}_i) \in \partial \mathcal{S},  \\
\quad  J &= \begin{bmatrix} \mathbi{V}^{\mathrm{raw}}_1 & \mathbi{V}^{\mathrm{raw}}_2 & \cdots & \mathbi{V}^{\mathrm{raw}}_{N_{\mathrm{impulses}}} \end{bmatrix}^\top, \\
\quad i &= 1, \ldots ,N_{\mathrm{impulses}},~~I_1 > I_2 > \cdots.
\end{split}
\end{equation}

Each impulse $\mathbi{V}^{\mathrm{raw}}_i$ consists of position $\mathbi{p}_i$, direction $\mathbi{d}_i$, and magnitude of strength $I_i$, and $\partial \mathcal{S}$ denotes the surfaces of the target mesh. Each impulse $\mathbi{V}^{\mathrm{raw}}_i$ is located on the element of triangle mesh with mesh area $A_i$ and the mesh normal $\mathbi{n}_i$. Also, the $\mathbi{V}^{\mathrm{raw}}_1$ denotes the principle impulse with the maximum strength $I_1$. The impulses $\mathbi{V}^{\mathrm{raw}}_i$ are ordered within $J$ by descending order of $I_i$.
$N_{\mathrm{impulses}}$ denotes the total number of impulses captured during the collision.

Regarding the BEM simulation \cite{hahn2016fast} uses an impulse threshold in a rigid-body system to determine whether to convert the impulses into stress for the BEM solver to process (as shown in Figure~\ref{fig:overview-new}, section (e)). In contrast, our framework employs the Maximum Normal Stress Criterion at local collision points \cite{Lubarda_Lubarda_2020} to decide whether to initiate the breaking process in the rigid-body system. This criterion is applied both during data generation and at run-time.
The critical fracture stress can be illustrated as follows:
\begin{equation}
\label{equ:critical-fracture-point}
\sigma_1 = \frac{I_1 \cdot \mathbi{d}_1 \cdot \mathbi{n}_1}{A_1 \cdot \Delta t } > \sigma_{0}.
\end{equation}

Since the $\sigma_{1}$ represents the maximum stress during a collision, we calculate whether this maximum principal stress exceeds the critical fracture stress. The critical fracture stress, denoted as $\sigma_0$, is determined using material parameters and is consistent with the value used in the data generation process described in \cite{hahn2016fast}. $\Delta t$ denotes the delta time since the collision occurred.
In Equation~\eqref{equ:critical-fracture-point}, we assume the Maximum Normal Stress Criterion is calculated on the local collision position of the principle impulse on the surface.

The framework initiates the breaking process if the conditional equation~\eqref{equ:critical-fracture-point} is satisfied. 
During the breaking process of the data generation period, we capture the impulses (as shown in Figure~\ref{fig:overview-new}, section (b)) and the fracture pattern $\mathcal{S}^{\prime}$ from the BEM fracture simulation process (as shown in Figure~\ref{fig:overview-new}, section (e)).
At run-time, our breaking process can be simplified as the function: 
\begin{equation}
\label{equ:learning-goal}
\mathcal{S}^{\prime} = \mathcal{G}(J)\
\end{equation}

$J$ is provided by the collision of the rigid-body system in Equation~\eqref{equ:impulse}. $\mathcal{S}^{\prime}$ denotes the fracture pattern we predict. $\mathcal{G}$ represents the generative model to predict fracture pattern, which will be detailed in Section~\ref{sec:network-architecture}.
In this paper, we evaluate simple collisions without scenes like pulling and twisting.
For the single collision scene in the main context of our paper, we use $J = \{\mathbi{V}^{\mathrm{raw}}_1\}$. For dual-collision scenes in the supplemental material, we use $J = \{\mathbi{V}^{\mathrm{raw}}_1, \mathbi{V}^{\mathrm{raw}}_2\}$. 
To address more complex multi-collision scenes, we need to increase the number of impulses and design specific scenarios to generate training data and retrain the generative model.

The supplemental material outlines the prerequisites and the algorithm's pseudocode to better understand the details of our proposed animation framework and existing crack propagation-based brittle fracture methods.

\begin{figure*}[tb]
  \centering
 \mbox{} \hfill
  \includegraphics[width=\linewidth]{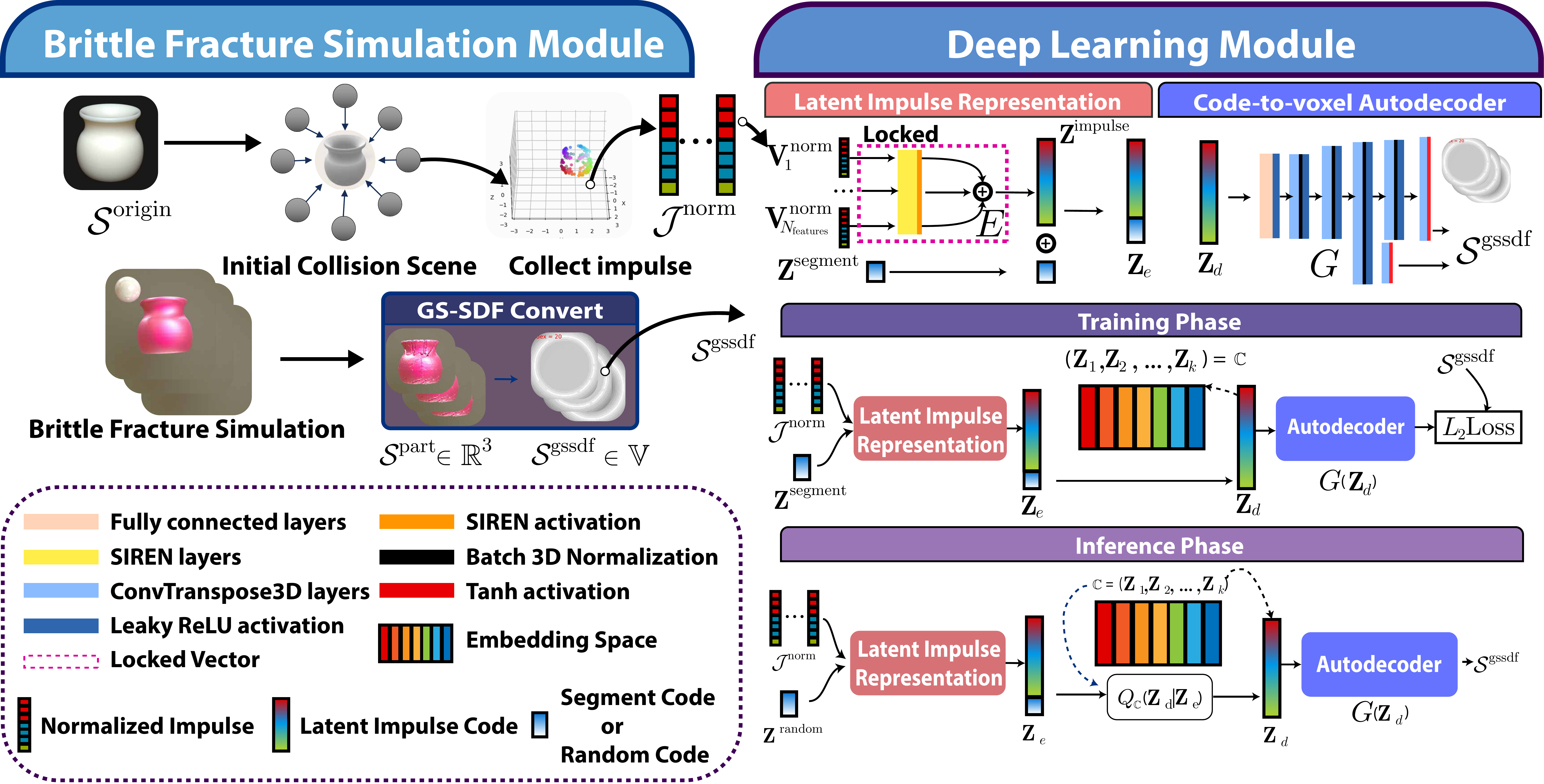}
  \vspace{-5mm}
 \hfill \mbox{}
  \caption{
    Our learning process consists of two modules: First, we design a scene to describe the collision and establish requirements to demonstrate our method's connection with brittle fracture physics. Consequently, we generate data by sampling the collisions between the ball and the target shape $\mathcal{S}^{\mathrm{origin}}$. Then, we collect the segmentation $\mathcal{S}^{\mathrm{part}}$ transformed into the GS-SDF $\mathcal{S}^{\mathrm{gssdf}}$.We pass the generated data through the Brittle Fracture Simulation Module and Deep Learning Module. The Deep Learning Module trains a code-to-voxel autodecoder ${G}$ with an embedding space $\mathbbm{c}$.}
  \label{fig:learning-process}%
\vspace{-5mm}
\end{figure*}

\section{Learning Process}
\label{sec:learning-process}

In the learning process shown in Figure~\ref{fig:learning-process}, we first adopt the set of breakable shapes $\mathcal{T}=\{t_1, t_2 \ldots , t_{N_{\mathrm{target}}}\}$ from the Thingi10k dataset \cite{Thingi10K}, where $N_{\mathrm{target}}$ is the number of breakable targets for training the generative model.
For each target $t \in \mathcal{T}$, we define the original shape $\mathcal{S}_t^{\mathrm{origin}}$ as the target shape.
In the brittle fracture simulation module, we apply the method by Hahn and Wojtan \cite{hahn2016fast} to create fractured shapes for training data, as detailed in Section \ref{sec:prerequisite} and \ref{sec:create-learning-data}.
We use these data to train a deep learning model to achieve Equation~\eqref{equ:learning-goal}, denoted as generator $\mathcal{G}$ with networks and parameters of $\{ E, G, \mathbbm{c} \}$, which includes an encoder $E$ with a SIREN layer for encoding latent impulse codes, an autodecoder $G$, and an embedding space $\mathbbm{c}$. We detail the network architecture in Section~\ref{sec:network-architecture}.

\subsection{Creating Learning Data}
\label{sec:create-learning-data}

The potential data format for the deep learning training input during collisions includes impulse information, external forces acting on the mesh surface, and kinetic energy before and after the collision.
In rigid-body engines like Bullet Physics (as discussed in Section~\ref{sec:prerequisite}), impulse information is used to represent the movement state before and after a collision. Therefore, we input impulse information into generative models to simulate energy transfer in collisions. The physics engine temporarily converts the kinetic energy of colliding objects into impulse information at the moment of collision. This information is then used to adjust the objects' velocities post-collision and to determine stress information in the BEM method.

We design a collision scene between a breakable target and an unbreakable sphere to simulate a single collision scene as described in Section \ref{sec:prerequisite}.
We assume that the target shape in a zero-gravity scene will be destroyed.
We position the target shape at the centre of a spherical space in such a scene. We then randomly shoot balls toward the centre from sampled positions on the surface of the spherical space. When a ball collides with the target shape, we store the impulse information occurring on the shape's surface as $J^{\mathrm{raw}}$. We also store the shape generated by the destruction simulation as $\mathcal{S}^{\mathrm{part}}$, and convert it to a \emph{Geometrically-Segmented Signed Distance Function} (GS-SDF) format $\mathcal{S}^{\mathrm{gssdf}}$, discussed later in Section~\ref{sec:gs-sdf}.

Using the brittle fracture simulation module, we conduct a large number of simulations to create a training dataset for target $ t \in \mathcal{T}$.
Each data pair consists of a collection of normalized impulse information vectors $J^{\mathrm{norm}}=\{\mathbi{V}^{\mathrm{norm}}_1, \ldots, \mathbi{V}^{\mathrm{norm}}_{N_{features}}\}$ and a GS-SDF $\mathcal{S}^{\mathrm{gssdf}}$.
The seven-dimensional impulse information vector 
$\mathbi{V}^{\mathrm{raw}} = [$\mathbi{p}$\: ${\mathbi{d}}$\: I]$ comes from the rigid-body dynamics in Equations~\eqref{equ:impulse} and \eqref{equ:learning-goal}, which
contains the position ($\mathbi{p} \in \mathbb{R}^{3}$), direction ($\mathbi{d} \in \mathbb{R}^{3}$), and scalar value ($I \in \mathbb{R}$) of each impulse. All impulse information is normalized into $\mathbi{V}^{\mathrm{norm}}$ within $[-1, 1]^{7}$. Note that we record the maximum strength of $I_{\mathrm{max}}$ and normalize $I$ by $\frac{2I}{I_{\mathrm{max}}}-1$.

To ensure consistency in the size of the target shapes $\{\mathcal{S}_t^{\mathrm{origin}} \mid t \in \mathcal{T} \}$ across the brittle fracture simulation module, the deep learning module, and the run-time process, all shapes are normalized in both the Euclidean space $\mathbb{R}^3$ and the voxel space $\mathbb{V}$. This normalization ensures that the longest length $l_\mathrm{max}$ of the bounding box equals 2.
In the deep learning module, a voxel space $\mathbb{V}$ of $[0,r]^{3}$ is created, with the centre of the object set to $(\frac{r}{2},\frac{r}{2},\frac{r}{2})$ in the voxel space $\mathbb{V}$. The Euclidean space $\mathbb{R}^3$ from $(-1,-1,-1)$ to $(1,1,1)$ corresponds to the voxel space $\mathbb{V}$ from $(0,0,0)$ to $(r,r,r)$, where $r$ denotes the resolution of the voxel space $\mathbb{V}$. 
In the experiments described later in this paper, we use $r=128$ or $256$.

\subsection{Network Architecture and Training Methodology}
\label{sec:network-architecture}

The objective of the learning process, as shown in Figure~\ref{fig:learning-process}, is to train a deep learning generative model $\mathcal{G}$ with networks and parameters of $\{ E, G, \mathbbm{c} \}$, learning a generative model $\mathcal{G}: J^{\mathrm{norm}} \mapsto \mathcal{S}^{\mathrm{gssdf}}$ from the collection of normalized impulse information vectors $J^{\mathrm{norm}}$ to a fracture pattern $\mathcal{S}^{\mathrm{gssdf}}$.
Our generative model $\mathcal{G}$ is specifically tailored to the dataset.

In this subsection, we first introduce the input representation in Section~\ref{sec:latent-impulse-representation} and the output representation in Section~\ref{sec:gs-sdf}.
Then, by introducing the concept of neural discrete representation learning in Section~\ref{sec:neural-discrete-learning}, we show how we process the input and output representations during the training phase in Section~\ref{sec:objective} and during the inference phase in Section~\ref{sec:inference}.

\subsubsection{Latent Impulse Representation}
\label{sec:latent-impulse-representation}

\begin{figure}[tb]
  \centering
  \includegraphics[width=\linewidth]{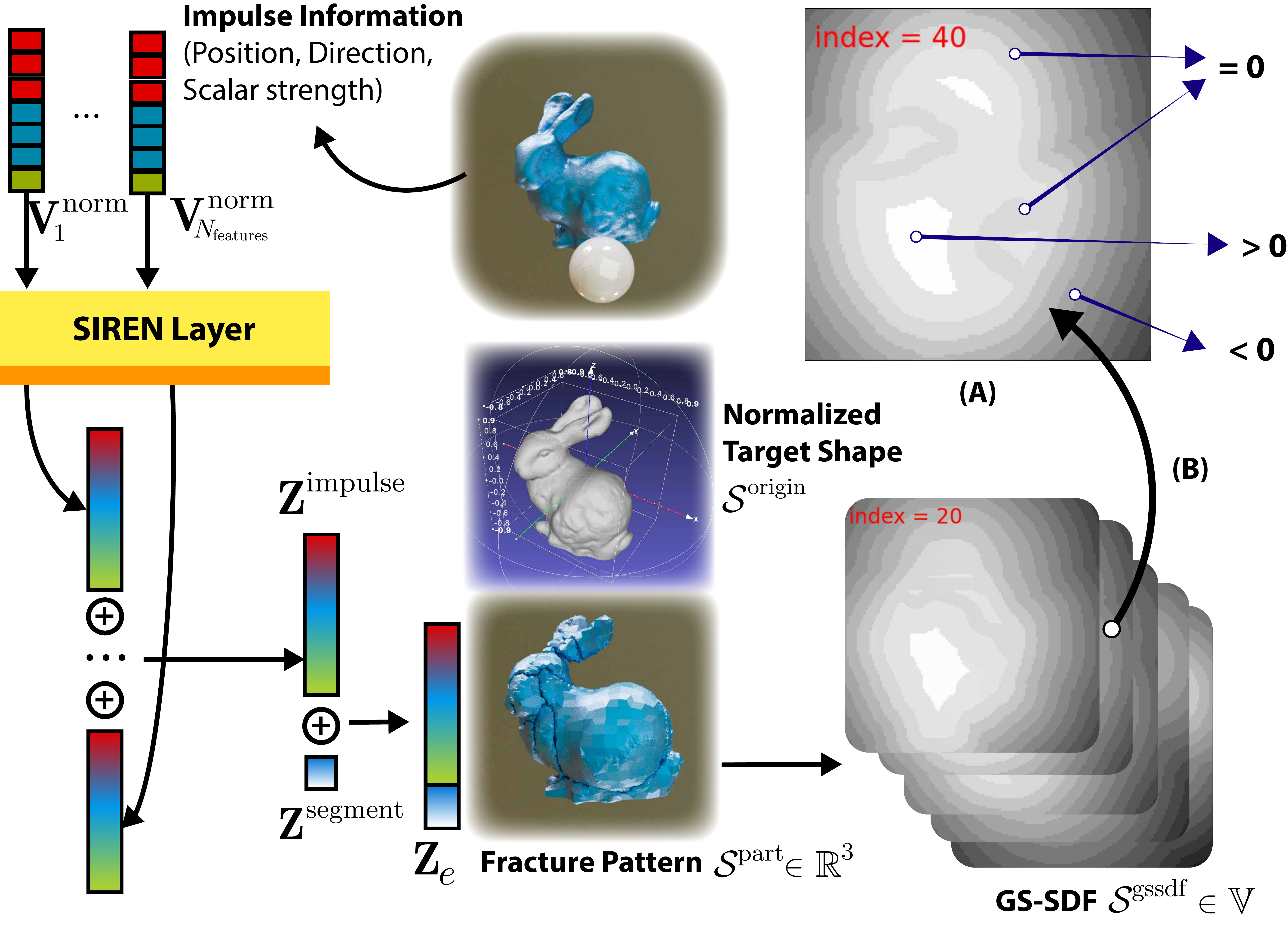}
  \vspace{-5mm}
  \caption{\label{fig:dataformat} 
  Design of latent impulse representation $\mathbi{Z}_e$ and GS-SDF $\mathcal{S}^{\mathrm{gssdf}}$. (A) A 2D Slice of GS-SDF, (B) 2D slice in depth 40th.}
  \vspace{-5mm}
\end{figure}

The latent impulse representation $\mathbi{Z}_e=\{\mathbi{Z}^{\mathrm{impulse}}, \mathbi{Z}^{\mathrm{segment}}\}$ concatenates the continuous latent impulse code $\mathbi{Z}^{\mathrm{impulse}}$ and the discrete segment code $\mathbi{Z}^{\mathrm{segment}}$, which follows a normal distribution.
Using the mapping $E : J^{\mathrm{norm}} \mapsto \mathbi{Z}^{\mathrm{impulse}}$, we obtain $\mathbi{Z}^{\mathrm{impulse}}$ through a higher-dimensional pre-trained encoding.
For each element $\mathbi{V}^{\mathrm{norm}}_i$ in $J^{\mathrm{norm}}$, we process it through a SIREN layer $E_{\mathrm{SIREN}}$ and apply a concatenation. As the result, $\mathbi{Z}^{\mathrm{impulse}} = (E_{\mathrm{SIREN}}(\mathbi{V}^{\mathrm{norm}}_1) \oplus \ldots \oplus E_{\mathrm{SIREN}}(\mathbi{V}^{\mathrm{norm}}_{N_{\mathrm{features}}}))$.

As shown in Figures~\ref{fig:learning-process} and \ref{fig:dataformat}, the normalized impulse code $\mathbi{V}^{\mathrm{norm}}_i \in [-1, 1]^7$ is a seven-dimensional vector, the latent impulse code $\mathbi{Z}^{\mathrm{impulse}} \in [-1, 1]^{N_{\mathrm{impulse}}^{\mathrm{dim}}}$, and the segment code $\mathbi{Z}^{\mathrm{segment}}$ or random code $\mathbi{Z}^{\mathrm{random}} \in [-1, 1]^{N_{\mathrm{segment}}^{\mathrm{dim}}}$ are vectors with different dimensions. 
The dimensions of $N_{\mathrm{impulse}}^{\mathrm{dim}}$ and $N_{\mathrm{segment}}^{\mathrm{dim}}$ need to be set as the suitable network parameter tailored to the training dataset, considering the number of fragments. We describe the details of the network parameters in Section~\ref{sec:exp-dataset}.

As illustrated in Figure~\ref{fig:learning-process}, our latent impulse representation differs between the training and inference phases. During inference, we use a random discrete normal distribution code $\mathbi{Z}^{\mathrm{random}}$ as the segment code $\mathbi{Z}^{\mathrm{segment}}$. It is important to note that $\mathbi{Z}^{\mathrm{random}}$ and $\mathbi{Z}^{\mathrm{segment}}$ share the same data structure and initialization method, with a random normal distribution.

Inspired by the autodecoder of DeepSDF \cite{park2019deepsdf} and the masking technique in the general generative networks \cite{yenduri2024gpt}, our segment code $\mathbi{Z}^{\mathrm{segment}}$ represents one segmented pattern under the same conditions of impulse and original target shape.
Our segment code is stored and backpropagated during training, whereas $\mathbi{Z}^{\mathrm{random}}$, used during inference, is randomly generated. 
This random generation aims to predict a possible encoded segment code based on the distribution of segment codes trained under the condition of $\mathbi{Z}^{\mathrm{impulse}}$.
This implies that the latent impulse code corresponds to the continuous latent space of the impulse condition. While the segment code represents the discrete latent space of multiple potential fracture patterns under similar impulse conditions. Consequently, our latent impulse representation is tailored explicitly for generating chaotic destruction fracture patterns.
In the supplemental material, we provide additional details and discuss the justification of segment code design in the ablation study.

Section~\ref{sec:inference} and Figure~\ref{fig:training-process} will further explain how we utilise the random code in our latent impulse representation to predict possible fractured shapes.

\subsubsection{Geometrically Segmented Signed Distance Function (GS-SDF)}
\label{sec:gs-sdf}

Following many experiences in 2D and 3D deep learning research, the potential data format of fractured shapes data in deep learning includes truncated signed distance fields (TSDF), unsigned distance fields (USDF), and masks.
Potentially tailored networks include the Convolutional Neural Network (CNN) and the Multilayer Perceptron (MLP) Neural Network. 
Even though USDF and masks were potential candidates for representing 3D fractures, we decided against using masks due to their integration failures with morphological segmentation, discussed later in Section~\ref{sec:caged-sdf}. Similarly, USDF was not adopted because it could not be integrated with caged-SDF segmentation, also discussed later in Section~\ref{sec:caged-sdf}. Early in our research, we concluded that the MLP network is unsuitable for our generative geometric segmentation task. This is because it failed to learn the adjacent fractured surfaces of multiple objects among all 3D representation candidates discussed above, trained by the autodecoder, and could only reproduce large fragments during the inference phase with DeepSDF \cite{park2019deepsdf}. We include the results of the inference phase comparisons between implicit and voxel-based representations in the supplemental material.
We also determined that MeshCNN and GraphNet are unsuitable for our generative geometric segmentation because the number of segments and the number of faces in these segment meshes vary with each fracture, and these networks cannot accommodate such variability.

Alternatively, in our approach, we define a 3D \emph{geometrically-segmented signed distance function} (GS-SDF) $\mathcal{S}^{\mathrm{gssdf}}$ in a real coordinate space $\mathbb{R}^3$.
The only difference between GS-SDF and the general signed distance function (pure SDF and TSDF) is that GS-SDF contains multiple objects in one shape.
As shown in Figure~\ref{fig:dataformat}, the distance field inside the shape is positive, while the values outside the shape are negative, and the surface and the fracture surface are nearly zero.

In the voxel space for learning, we define $D(\mathbi{p})$ as the shortest Euclidean distance from the coordinate position $\mathbi{p}$ to all boundary surfaces of the divided shape $\{S_r \mid S_r \subset \mathcal{S}^{\mathrm{part}}, r = 1, \ldots, N_{\mathrm{regions}}\}$, where $N_{\mathrm{regions}}$ is the number of region divisions. In this case, the unsigned distance field (USDF) inside the learning target shape can be defined as: $D(\mathbi{p})=\min(D_1(\mathbi{p}), D_2(\mathbi{p}), \ldots ,D_{N_{\mathrm{regions}}}(\mathbi{p}))$ and $D_{\mathrm{origin}}(\mathbi{p})$ can be specially treated as the Euclidean distance from the surface of the target shape $\mathcal{S}^{\mathrm{origin}}$. By integrating external and internal voxel space, we can define the GS-SDF $\mathcal{S}^{\mathrm{gsdf}}$ as $\{ f^{\mathrm{gssdf}}(\mathbi{p}) \mid \mathbi{p} \in \mathbb{R}^3 \}$ as follows (see Figure~\ref{fig:dataformat} upper right):
\begin{equation}
\label{equ:gssdf}
    \begin{split}
        f^{\mathrm{gssdf}}(\mathbi{p}) = 
\begin{cases}
    D(\mathbi{p}),& \mathrm{if }~~\mathbi{p} \in \mathcal{S}^{\mathrm{part}}\\
    -D_{\mathrm{origin}}(\mathbi{p}).  & \mathrm{otherwise}
\end{cases}
    \end{split}
\end{equation}

\subsubsection{Concept of Neural Discrete Representation Learning}
\label{sec:neural-discrete-learning}

Learning brittle fracture patterns faces the challenge of mapping a continuous representation of $J^{\mathrm{norm}}$ to a discrete representation of $\mathcal{S}^{\mathrm{gssdf}}$.
In brittle fracture, two different fracture patterns with large Euclidean distances may arise from similar collision conditionsdue to the chaotic nature of the process.
The distribution of fracture pattern representation is sparse and discrete, making it difficult to use a continuous representation of the impulse information vectors $J^{\mathrm{norm}}$ to train a generator for predicting a discrete output representation of $\mathcal{S}^{\mathrm{gssdf}}$. It is challenging to map a single continuous representation to multiple discrete representations.

To address these challenges, we introduce an embedding space $\mathbbm{c}$, a discrete latent code mapping technique called \emph{vector quantisation} (VQ) $Q_\mathbbm{c}$ in the neural discrete representation learning \cite{van2017neural}, and the code-to-voxel autodecoder $G$ inspired by \cite{park2019deepsdf} and \cite{wu2016learning}.
Neural discrete representation learning, known as VQ-VAE, separates the encoder and decoder combination found in a typical VAE network and employs an autoregressive decoder to generate an embedding space. This space is represented as a dictionary of discrete latent codes. Using these discrete latent codes, the encoder maps the input data to each discrete latent code through a process known as vector quantisation, which involves searching nearby mappings.

In this paper, we adopt the concept of the embedding space and discrete latent codes from \cite{van2017neural} as the embedding space $\mathbbm{c}$ and discrete latent code  $\mathbi{Z}_k$ in Figure~\ref{fig:learning-process}. 
Since the VQ-VAE splits the autoencoder into the distinct encoder and autoregressive generator components, the training concept of the autoregressive generator is the same as the autodecoder in DeepSDF \cite{park2019deepsdf}.
This approach allows us to train the embedding space with a decoder-only framework and use our latent impulse representation as an encoder. 
Additionally, inspired by \cite{wu2016learning}, we have designed a customized code-to-voxel autodecoder to substitute for the autodecoder in both DeepSDF and VQ-VAE, specifically for processing 3D voxel data.

By collecting latent vectors trained by the autodecoder in Section~\ref{sec:generator} we can build a dictionary of discrete latent codes called embedding space $\mathbbm{c}$, which represents the discrete embedding space of all fracture patterns.
In the training and inference phases, detailed in Sections~\ref{sec:objective} and \ref{sec:inference}, we successfully map the continuous representation of $J^{\mathrm{norm}}$ to a discrete representation of $\mathcal{S}^{\mathrm{gssdf}}$. This approach avoids direct training of the decoder using a continuous representation. 
Additionally, by swapping the random code $\mathbi{Z}^{\mathrm{random}}$ with the segment code $\mathbi{Z}^{\mathrm{segment}}$ during the inference phase, as outlined in Sections~\ref{sec:latent-impulse-representation} and \ref{sec:inference}, we address the challenge of mapping a single input to multiple possible outputs.

\subsubsection{Code-to-Voxel Autodecoder}
\label{sec:generator}

In our approach, we define the autodecoder as ${G}:\mathbi{Z}_d \mapsto \mathcal{S}^{\mathrm{gssdf}}$ in Figure~\ref{fig:learning-process}.
During the training phase, we use the discrete latent code $\mathbi{Z}_d$ as the latent impulse representation $\mathbi{Z}_e$. The latent impulse representation $\mathbi{Z}_e$ will be used to train the generator $G$, which maps $(E(J^{\mathrm{norm}}), \mathbi{Z}^{\mathrm{segment}})$ to $\mathcal{S}^{\mathrm{gssdf}}$. 
While the mapping $E$ does not undergo backpropagation, the segment code will be encoded by the autodecoder $G$. At the end of each training epoch, we sample the discrete latent code in $\mathbbm{c}$. 

In the inference phase, we obtain the discrete latent code $\mathbi{Z}_d$ from the elements of embedding space  $\mathbbm{c}$, which means ${G}$ will be regarded as a decoder for discrete latent code. $\mathbi{Z}_d$ shares the same dimension with $\mathbi{Z}_e$ and the element of embedding space $\mathbbm{c}$.

\subsubsection{Training Phase in Neural Discrete Representation Learning}
\label{sec:objective}

\begin{figure}[tb]
  \centering
  \includegraphics[width=\linewidth]{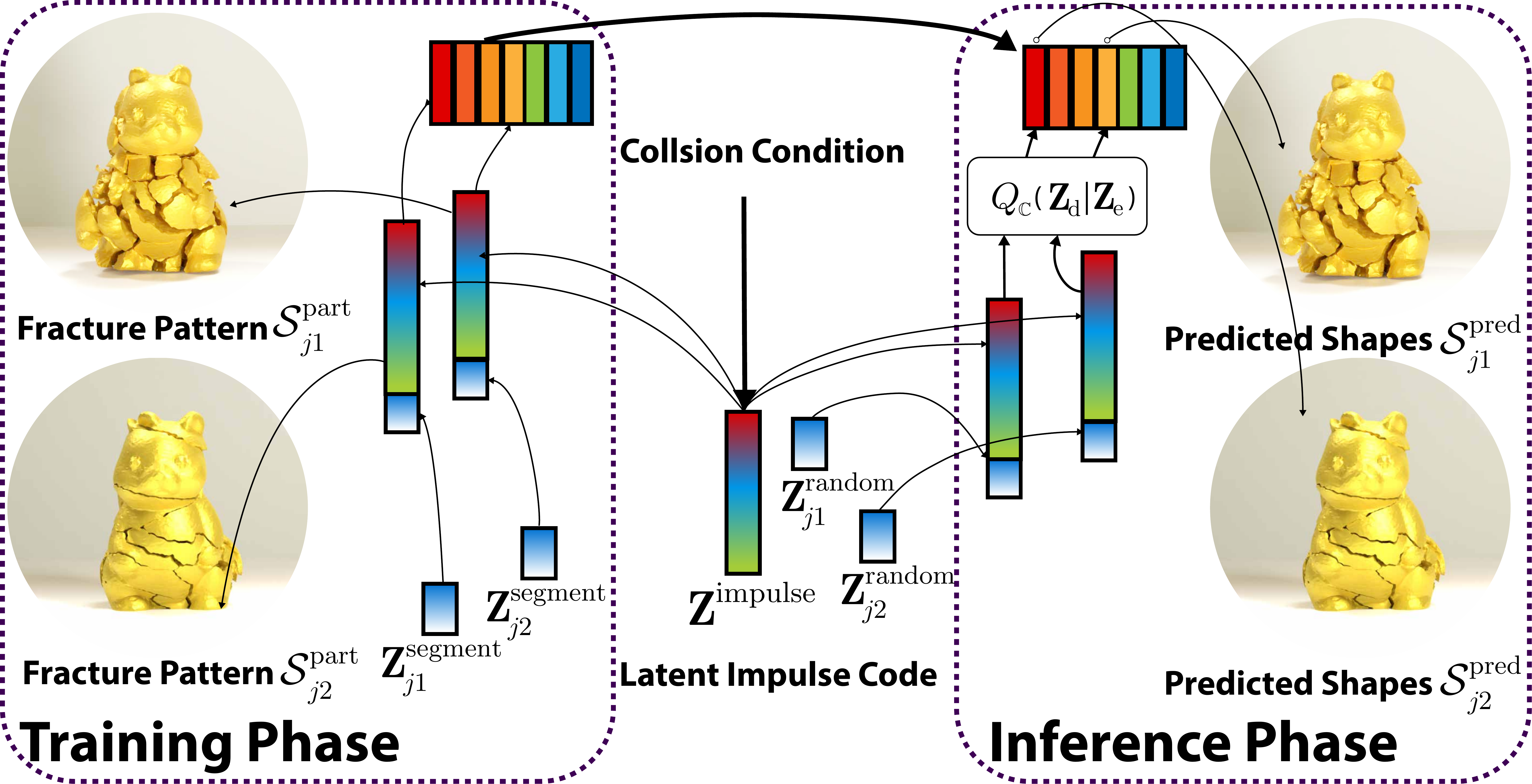}
  \vspace{-5mm}
  \caption{
  Illustration of the relationship and differences between the segment code $\mathbi{Z}^{\mathrm{segment}}$ and the random code $\mathbi{Z}^{\mathrm{random}}$, as defined in Section~\ref{sec:latent-impulse-representation}. The segment code $\mathbi{Z}^{\mathrm{segment}}$ is trained during the training phase, while the random code $\mathbi{Z}^{\mathrm{random}}$ is used during the inference phase:
  During the training phase, $\mathbi{Z}^{\mathrm{segment}}$ is treated as $\mathbi{Z}^{\mathrm{segment}}_i$, which is initialized with a normal distribution tailored to $\mathcal{S}^{\mathrm{part}}_i$. However, at run-time, the segment code $\mathbi{Z}^{\mathrm{segment}}$ is treated as a random noise $\mathbi{Z}^{\mathrm{random}}_i$. By searching the closest latent vector in embedding space with different $\mathbi{Z}^{\mathrm{random}}_{j1}$ or $\mathbi{Z}^{\mathrm{random}}_{j2}$, we can obtain completely different but reasonable shapes of $\mathcal{S}^{\mathrm{part}}_{j1}$ or $\mathcal{S}^{\mathrm{part}}_{j2}$ under the same condition of $\mathbi{Z}^{\mathrm{impulse}}$. 
  }
  \vspace{-5mm}
  \label{fig:training-process}
\end{figure}

Notably, according to the training scheme described in Section~\ref{sec:neural-discrete-learning} and \ref{sec:generator}, by substituting the $\mathbi{Z}_e$ into $\mathbi{Z}_d$ in the training phase, our autodecoder network $G$ can be trained as the mapping $G: (E(J^{\mathrm{norm}}), \mathbi{Z}^{\mathrm{segment}}) \mapsto \mathcal{S}^{\mathrm{gssdf}}$, which means we can simply learn the mapping according to this objective:
\begin{equation}
\label{equ:objective-1}
\begin{split}
\mathcal{L}_{L_2}(G, E) =& \mathbb{E} [\|\mathcal{S}^{\mathrm{gssdf}} - G(E(J^{\mathrm{norm}}), \mathbi{Z}^{\mathrm{segment}})\|_{2}],
\end{split}
\end{equation}
\begin{equation}
\label{equ:objective-4}
G^{*} = \mathrm{arg} \; \mathop{\min}_{G} \;[ \mathcal{L}_{L_2}(G, E)].
\end{equation}
Note that with this objective, $E$ does not undergo backpropagation, ensuring that the distribution of $\mathbi{Z}^{\mathrm{impulse}}$ remains unchanged during training. This stability allows us to successfully encode the segment code $\mathbi{Z}^{\mathrm{segment}}$ using the autodecoder $G$.
We assign each shape $\mathcal{S}^{\mathrm{gssdf}}_i$ an initial random normal distribution code $\mathbi{Z}^{\mathrm{segment}}_i$, indicating that backpropagation will alter the distribution of the code $\mathbi{Z}^{\mathrm{segment}}_i$ associated with the segment of $\mathcal{S}^{\mathrm{gssdf}}_i$, as illustrated in the left part of Figure~\ref{fig:training-process}. The segment code $\mathbi{Z}^{\mathrm{segment}}_i$ is stored in $G$ and serves as the shape code for $\mathcal{S}^{\mathrm{gssdf}}_i$ during the training phase.

After completing training in each epoch, we randomly sample the concatenation of $\{\mathbi{Z}^{\mathrm{impulse}}_i, \mathbi{Z}^{\mathrm{segment}}_i \}$ as $\mathbi{Z}_{k}$ and generate the embedding space $\mathbbm{c}$ as described in Equation~\eqref{equ:embedding space}. This embedding space $\mathbbm{c}$ is then stored as a parameter of the generative model $\mathcal{G}$ and is used during the run-time process. The dimension of $\mathbi{Z}_k$ matches that of $\mathbi{Z}_e$ and $\mathbi{Z}_d$:
\begin{equation}
\label{equ:embedding space}
\begin{split}
\mathbbm{c} &= (\mathbi{Z}_1, \mathbi{Z}_2,\ldots, \mathbi{Z}_{N^{\mathrm{space}}}),  \;\; k = 1, 2, \ldots , N^{\mathrm{space}}, \\
\end{split}
\end{equation}
where $N^{\mathrm{space}}$ denotes the number of discrete latent codes stored in $\mathbbm{c}$.

\begin{figure*}[tb]
  \centering
  \mbox{} \hfill
  \includegraphics[width=\linewidth]{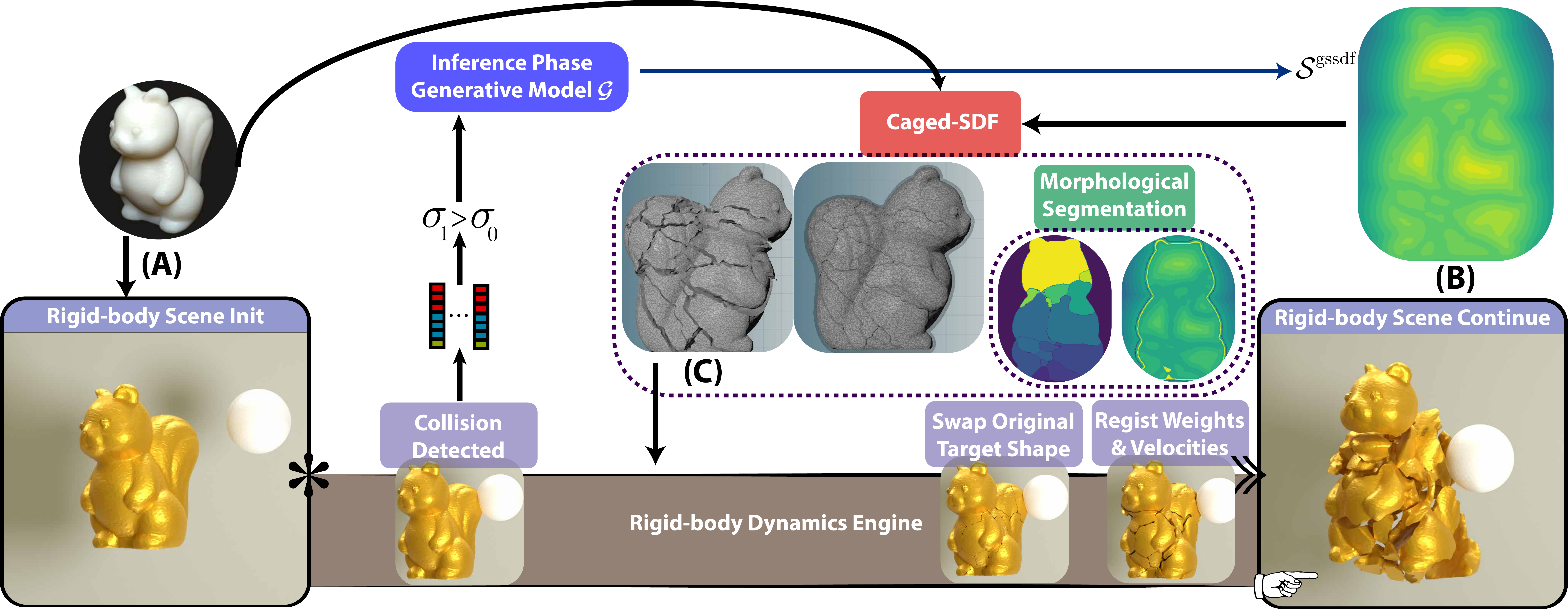}
  \vspace{-5mm}
  \hfill \mbox{}
  \caption{
    Flowchart of the run-time process: Once a collision occurs and the impulse exceeds the threshold, we input $J^{\mathrm{norm}}$ to the generative model $\mathcal{G}$. After preprocessing the prediction of voxel data $\mathcal{S}^{\mathrm{gssdf}}$ and applying caged-SDF segmentation, we assign weights and velocities to the fragmented shapes $\mathcal{S}^{\mathrm{part}}$. By replacing the old rigid body $\mathcal{S}^{\mathrm{origin}}$ with new rigid bodies $\mathcal{S}^{\mathrm{part}}$, we continue the rigid body simulation. (A) Original mesh; (B) GS-SDF $\mathcal{S}^{\mathrm{gssdf}}$; (C) Results of caged-SDF segmentation $\mathcal{S}^{\mathrm{part}}$.
    }
  \vspace{-5mm}
  \label{fig:run-time-process}%
\end{figure*}

\subsubsection{Inference Phase in Neural Discrete Representation Learning}
\label{sec:inference}

During the inference phase, we generate a new random normal distribution code $\mathbi{Z}^{\mathrm{random}}$ as the segment code and receive a collection of normalized impulse code $J^{\mathrm{norm}}$ from the run-time process, as depicted in the right part of Figure~\ref{fig:training-process}.
With the embedding space $\mathbbm{c}$ and the encoded latent vector $\mathbi{Z}_e = (E(J^{\mathrm{norm}}), \mathbi{Z}^{\mathrm{random}})$ during this phase, we generate the discrete latent vector $\mathbi{Z}_d$ using vector quantization $Q\mathbbm{c}(\mathbi{Z}_d \mid \mathbi{Z}_e)$ based on the embedding space $\mathbbm{c}$.
In addition, because $\mathbbm{c}$ is updated in the generative model $\mathcal{G}$ during the training phase in each epoch, we use the same version of $\mathbbm{c}$ as the generator  $G$ in the same epoch.

\begin{align}
\label{equ:quantilize}
    \begin{split}
       &  Q_\mathbbm{c}(\mathbi{Z}_d \mid \mathbi{Z}_e) = 
\begin{cases}
    1, & \mathrm{for} \;\;\;  k = \mathrm{argmin}_j \| \mathbi{Z}_e - \mathbi{Z}_j \|_2, \\
    0, & \mathrm{otherwise},
\end{cases}
    \end{split} \\
\label{equ:zd}
      &  \mathbi{Z}_d = \mathbi{Z}_k, \;\;\; \mathrm{where} \;\;\;  k = \mathrm{argmin}_j \| \mathbi{Z}_e - \mathbi{Z}_j \|_2.
\end{align}
With Equations~\eqref{equ:quantilize} and \eqref{equ:zd}, we can obtain $\mathbi{Z}_d$, which is derived from the closest discrete latent code $\mathbi{Z}_{k}$ in the embedding space $\mathbbm{c}$, identified through nearby searching using Euclidean distance.

As illustrated in the right part of Figure~\ref{fig:training-process}, by utilizing different random codes $\mathbi{Z}^{\mathrm{random}}_{j1}$ and $\mathbi{Z}^{\mathrm{random}}_{j2}$, combined with the same $\mathbi{Z}^{\mathrm{impulse}}$, our inference phase can generate two distinct yet plausible predicted fractured shapes under the same collision condition. 

Finally, we obtain the predicted fracture pattern $\mathcal{S}^{\mathrm{gssdf}} = G(\mathbi{Z}_\mathrm{d})$ with the mapping $J^{\mathrm{norm}} \mapsto \mathcal{S}^{\mathrm{gssdf}}$ during the inference phase.
In addition, further discussion about the loss function is provided in the supplement material.


\section{Run-time Process}
\label{sec:run-time-process}

During the run-time process depicted in Figure~\ref{fig:run-time-process}, a physics engine such as Bullet Physics carries out rigid body simulations. When a collision with the target shape $\mathcal{S}^{\mathrm{origin}}$ occurs, it captures the impulse data $J^{\mathrm{raw}}$ in Equation~\eqref{equ:impulse}. The impulse data tests the critical fracture stress in Equation~\eqref{equ:critical-fracture-point}. Once the test passes, the impulse data $J^{\mathrm{raw}}$ will be transmitted to the conditional generative model $\mathcal{G}$. 
This model predicts a fracture pattern $\mathcal{S}^{\mathrm{gssdf}}$ in the inference phase, which is promptly segmented into shapes $\mathcal{S}^{\mathrm{part}}$ using caged-SDF segmentation. The simulation then proceeds by substituting the original model $\mathcal{S}^{\mathrm{origin}}$ with these newly formed segmented shapes $\mathcal{S}^{\mathrm{part}}$.

\subsection{Collision in Pre-processing Module and Prediction Module}
\label{sec:preprocess}

If a collision occurs, all impulses of the target shape $\mathcal{S}^{\mathrm{origin}}$ within that frame are collected. 
If the maximum principal stress calculated by Equation~\eqref{equ:critical-fracture-point} reaches or exceeds the critical fracture stress, raw impact information $J^{\mathrm{raw}}$ is extracted. 
Concurrently, the velocity and mass $\{ \mathbi{v}^{\mathrm{origin}}, m^{\mathrm{origin}} \}$ of the target shape $\mathcal{S}^{\mathrm{origin}}$ before the collision are recorded. The method for creating normalized impact information $J^{\mathrm{norm}}$ from $J^{\mathrm{raw}}$ is detailed in Section~\ref{sec:create-learning-data}. Subsequently, the fracture pattern $\mathcal{S}^{\mathrm{gssdf}}$ is predicted by the conditional generative model $\mathcal{G}$ under the condition $J^{\mathrm{norm}}$, as described in the inference phase in Section~\ref{sec:inference}.

\begin{figure}[tb]
  \centering
  \includegraphics[width=\linewidth]{figs/7-caged-sdf.pdf}
  \vspace{-5mm}
  \caption{
 Illustration of caged-SDF segmentation: (A) USDF of original mesh ${\mathcal{S}^\mathrm{origin}}$ (yellow); (B) GS-SDF $\mathcal{S}^{\mathrm{gssdf}}$; (C) GS-SDF with flexible cage ${\mathcal{S}^\mathrm{gssdf}}^{\prime}$ (black) and original mesh ${\mathcal{S}^\mathrm{origin}}$ (yellow); (D) Labeled shape regions $\mathcal{S}^{\mathrm{labels}}$ generated by \cite{morpholibj2016}; (E) Morphological segmentation; (F) Isosurface extraction algorithm; (A’) Original mesh $\mathcal{S}^{\mathrm{origin}}$; (C’) Result of caged-SDF segmentation $\mathcal{S}^{\mathrm{part}}$; (D’) Cage with fractured surfaces in a mesh $\mathcal{S}^\mathrm{cage}$.
  }
  \vspace{-5mm}
  \label{fig:caged-sdf-process}
\end{figure}

\subsection{Caged-SDF Segmentation}
\label{sec:caged-sdf}

Interpolating thin, fractured surfaces from wide gaps and providing a cage to preserve the external original mesh during cutting are common challenges in brittle fracture animation research \cite{brittlempm2022, breakinggood2023}. 
As illustrated in Figure~\ref{fig:caged-sdf-process}, we have developed a method called \emph{caged-SDF segmentation} to reconstruct the destruction patterns, denoted as $\mathcal{S}^{\mathrm{part}}$, while retaining the original external surface mesh $\mathcal{S}^{\mathrm{origin}}$.

The caged-SDF segmentation method aims to generate predicted internal fractured surfaces, which are enclosed by a thin, soft-wrapping cage denoted as $\mathcal{S}^\mathrm{cage}$ (D') in Figure~\ref{fig:caged-sdf-process}. This approach involves several Boolean set operations between the cage with fractured surfaces $\mathcal{S}^\mathrm{cage}$ (D') and the original mesh $\mathcal{S}^\mathrm{origin}$ (A') in Figure~\ref{fig:caged-sdf-process}.

It is important to use a flooding method like morphological segmentation (E) to generate the labelled regions because directly processing only Boolean set operations would generate unconnected and non-manifold surfaces. The flooding algorithm will help us create watertight and manifold fragment shapes connecting the original mesh and fractured surfaces.

As shown in Figure~\ref{fig:caged-sdf-process}, we begin by receiving the original mesh $\mathcal{S}^\mathrm{origin}$ (A') from the rigid-body system during runtime. After accepting the $\mathcal{S}^{\mathrm{gssdf}}$ prediction (B) from generative model $\mathcal{G}$, we then create a GS-SDF with a flexible cage, ${\mathcal{S}^\mathrm{gssdf}}^{\prime}$ (C). We do this by adding a small constant, $\epsilon = 0.03$, which alters the zero level-set in voxel space. Next, we perform morphological segmentation (E) on ${\mathcal{S}^\mathrm{gssdf}}^{\prime}$ (C), which results in labeled shape regions $\mathcal{S}^\mathrm{labels}$ in voxel space (D). Using the isosurface extraction algorithm (F), we create the mesh $\mathcal{S}^{\mathrm{cage}}$ shown in (D'). Finally, we apply Boolean set operations between this cage mesh and the original mesh, producing the segmented mesh $\mathcal{S}^{\mathrm{part}}$ (C').

\begin{figure*}[tb]
  \centering
  \mbox{} \hfill
  \includegraphics[width=\linewidth]{figs/exp-compare-1.pdf}
  \vspace{-5mm}
  \hfill \mbox{}
  \caption{
    Comparison between the simulation results, results by Sell\'{a}n et al., and fractured shapes results predicted by deep learning.
    Left to right: Input collision condition, Brittle fracture simulation results, Results of our method, Results by Sell\'{a}n et al. \cite{breakinggood2023}.
    Top to bottom row: Squirrel, Pot, Bunny, Base.
    Note that two random normal distribution codes generated our results, and we selected one, which is the process shown in the run-time process in Figure~\ref{fig:training-process}.
    For the results by Sell\'{a}n et al. \cite{breakinggood2023}, we computed the system with 6000 cages and from 5 to 60 modes and selected the best visualization of all the test cases. No collision conditions are contained in the learning process.
    }
    \vspace{-5mm}
    \label{fig:exp-compare-1}
\end{figure*}

\paragraph*{Morphological segmentation.} 
A common flooding algorithm-based approach is widely used for 3D instance segmentation and cell division in medical imaging research~\cite{wang2022novel}.
With an unsigned distance field (USDF) $\mathcal{S}^{\mathrm{usdf}}$ generated by multiplying the values of ${\mathcal{S}^\mathrm{gssdf}}^{\prime}$ less than 0 by -1, we perform stable flooding-based instance segmentation executed cell-by-cell by raising the threshold in 0.04 increments by using the method in \cite{morpholibj2016}. 
Note that, the zero level-set in ${\mathcal{S}^\mathrm{gssdf}}^{\prime}$ is the same as $-\epsilon$ level-set in $\mathcal{S}^\mathrm{gssdf}$.
If we use the flooding method for instant segmentation with ${\mathcal{S}^\mathrm{gssdf}}^{\prime}$, the flooding of each region will be stopped in the new edge of the cage, extracted by zero level-set in ${\mathcal{S}^\mathrm{gssdf}}^{\prime}$. Thus, the cage will be connected to internal fractured surfaces extracted by $\epsilon$ level-set in ${\mathcal{S}^\mathrm{gssdf}}^{\prime}$, which is an internal zero level-set in $\mathcal{S}^\mathrm{gssdf}$. 
All uniquely labelled shape regions $\mathcal{S}^{\mathrm{labels}}$ (D) are reconstructed into a 3D mesh using an isosurface extraction algorithm such as marching cubes, forming the mesh $\mathcal{S}^{\mathrm{cage}}$ shown in (D’) of Figure~\ref{fig:caged-sdf-process}.

\paragraph*{Boolean set operations.}
First, we perform a Boolean intersection operation between $\mathcal{S}^{\mathrm{origin}}$ (A') and $\mathcal{S}^{\mathrm{cage}}$ (D'). Subsequently, a Boolean union operation between the intersection result and $\mathcal{S}^{\mathrm{origin}}$ allows us to derive an internal fractured shape with the external original mesh $\mathcal{S}^{\mathrm{part}}$ (C'), as shown in the ``Mesh Boolean'' of Figure~\ref{fig:caged-sdf-process}. 

\subsection{Reconstruction and Post-processing}
\label{sec:post-process}

The mass of each fragment in $\mathcal{S}^{\mathrm{part}}$ is determined based on its size relative to the original shape $\mathcal{S}^{\mathrm{origin}}$. Using the velocity of the shape $\mathbi{v}^{\mathrm{origin}}$ stored before being replaced in Section~\ref{sec:preprocess}, the velocity $\{ \mathbi{v}_r \mid \mathbi{v}_r = \mathbi{v}^{\mathrm{origin}} \}$ is distributed evenly to the newly formed fragment shapes as rigid-body information.

In the final step, the original pre-destruction shape is removed. The reformed fragment shapes, each with its mass $m_r$ and velocity $\mathbi{v}_r$, are incorporated as rigid bodies into a physics engine. The destruction animation is then completed by computing subsequent frames in the rigid-body simulation.

\section{Experimental Results and Discussion}
\label{sec:experiment}

\subsection{Dataset and Implementation}

In all experiments, we used the shape data scanned by Thingi10K \cite{Thingi10K}. We selected four models: Pot (Thing ID:12120), Bunny (Thing ID:240197), Squirrel (Thing ID:11705), and Base (Thing ID:17204). We conducted both the learning and run-time processes for each target shape with a generative model tailored to one target shape individually. 
We also performed comparative experiments with the simulation results of Hahn and Wojtan \cite{hahn2016fast} and Sell\'{a}n et al. \cite{breakinggood2023}. 
For the quantitative comparison, we implemented the 3D centroidal Voronoi diagram (CVD) by following the method of Schvartzman et al. \cite{schvartzman2014fracture} with handmade Voronoi control points to compare the size distribution, surface normals, and size histogram.

\subsection{Creation of Learning Data and Deep Learning}
\label{sec:exp-dataset}

Using the brittle fracture simulation module described in Figure~\ref{fig:learning-process}, calculations were carried out over four days on four PCs with Ryzen 9 5950X CPUs. For each of the four learning target shapes, Pot, Bunny, Squirrel, and Base, we conducted 200-frame destruction experiments 300 times in the scene environment introduced in Section~\ref{sec:create-learning-data}.
To include impulses close to the critical fracture impulse, our experiments also involve cases without any breaking. As a result, after excluding the non-fracturing cases, we collected 250 sets of input and output data for each shape.

\begin{figure}[tb]
  \centering
      \includegraphics[width=\linewidth]{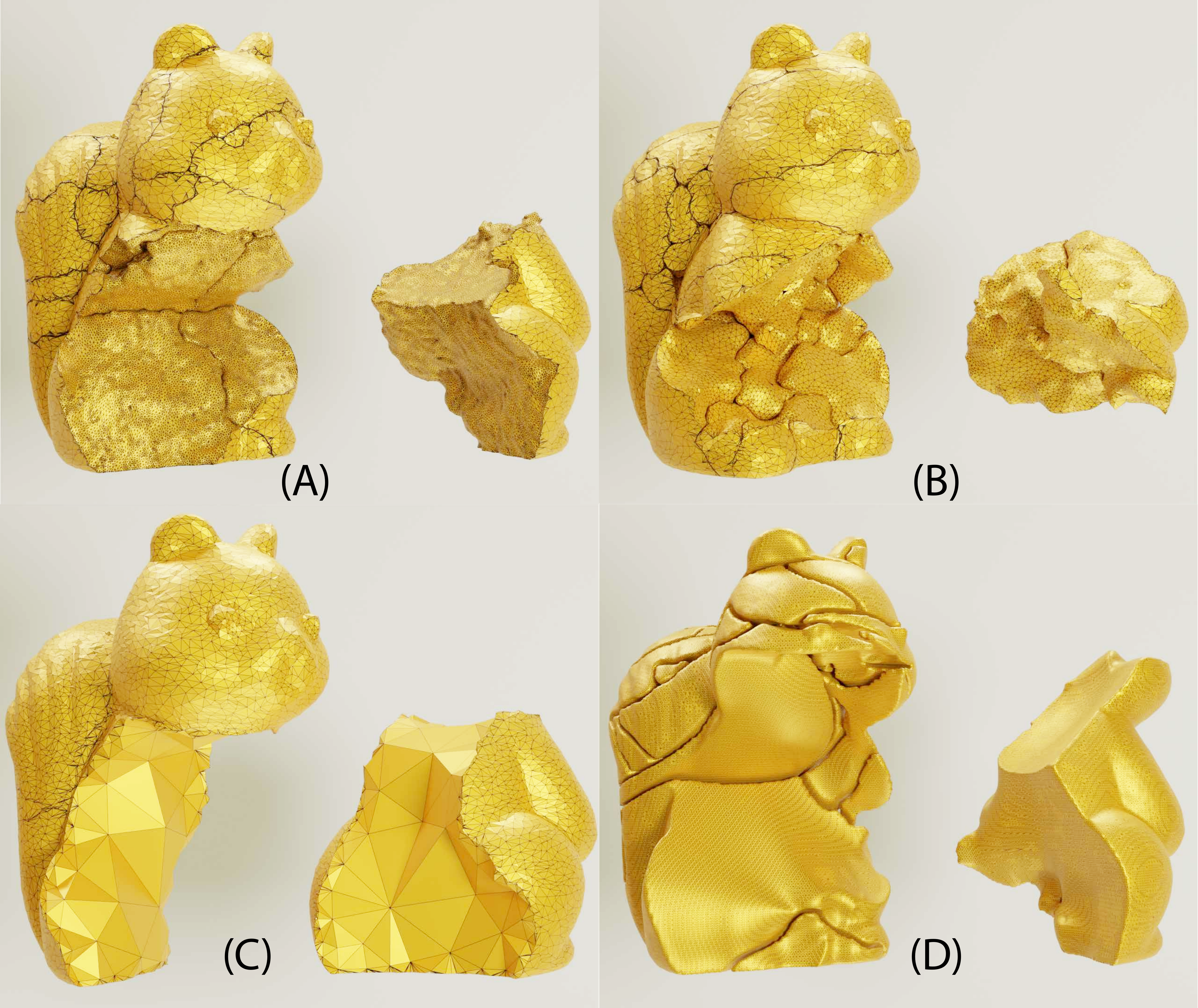}
 \vspace{-5mm}
  \caption{
  Comparison of fractured surfaces using our method at resolutions $r = 256$ (A) and $r = 128$ (B) for $\mathcal{S}^{\mathrm{gssdf}}$, Sellán et al.'s method (C), and BEM brittle fracture simulation (D):
  All outcomes were produced under similar collision conditions that were not part of the training dataset. Note that Sellán et al.'s results were computed using a resolution of $13{\small,}000$ cages with 10 modes. All images are rendered using both flat and wireframe shading.
  }
  \label{fig:exp-comparison-shape-2}
  \vspace{-5mm}
\end{figure}

We used 200 sets for training and 50 sets for testing, as shown in Figures~\ref{fig:exp-compare-1} and \ref{fig:exp-comparison-shape-2}. We set the dimension of $N_{\mathrm{impulse}}^{\mathrm{dim}}$ to 128, $N_{\mathrm{segment}}^{\mathrm{dim}}$ to 8, and $N^{\mathrm{space}}$ to 200.
Because of the initial single collision scene described in Section~\ref{sec:create-learning-data}, our experiments are all set with $N_{\mathrm{features}}$ to 1. The supplemental material will show the results of $N_{\mathrm{features}} = 2$.
In this study, we trained our networks at two resolutions $r$ of GS-SDF $\mathcal{S}^{\mathrm{gssdf}}$, specifically $r = 128$ and $r = 256$, to evaluate both the quality of the results and the computational time.
The results shown in Figure~\ref{fig:exp-compare-1} were obtained using the network configured with $r = 128$, while only the result labelled (A) in Figure~\ref{fig:exp-comparison-shape-2} used the network with $r = 256$.

The training time for all learning target shapes was 1000 epochs since we have few data compared with large datasets. 
We use the 1000th epoch for testing in Figure~\ref{fig:exp-compare-1}. 
For better stability verification of the model, we set the batch size as 1, which resulted in an average of 25.88 hours to complete each learning process. The model's parameters were updated with a learning rate of 0.003 for the first 400 epochs, followed by 0.00005 for the remaining 600 epochs.
The impulses of position, direction, and strength used in the testing (shown in Figures~\ref{fig:exp-compare-1} and \ref{fig:exp-comparison-shape-2}) were not included during the training phase.

In the supplemental material, we present the training loss and the training time with different epochs for the results in Figure \ref{fig:exp-compare-1} and discuss the suitable epoch for each training. 

\subsection{Comparison of Brittle Fracture Shape Prediction Results with Simulation Results}

Figure~\ref{fig:exp-compare-1} shows that our results are similar to brittle fracture simulation results regarding destruction patterns, global complexity, and fracture surface shapes. 
Sell\'{a}n et al.'s method needs to adjust the number of fragments by regenerating the pre-fractured pattern with different parameters of mode, as shown in the Squirrel and Pot examples in Figure~\ref{fig:exp-compare-1}.
Compared to Sell\'{a}n et al., our method occasionally generates fewer fractured shapes, but the patterns are closer to the result of brittle fracture simulation and tailored to specific collision conditions.

Figure~\ref{fig:exp-comparison-shape-2} illustrates that our fractured surfaces created by the network, even with the resolution of $r = 128$, can produce complex shapes and visually impactful internal fractured surfaces while preserving the external original mesh. 
Since the method of Hahn and Wojtan \cite{hahn2016fast} do not provide the option to preserve the external original mesh, we present a re-meshed version of the original surface in Figure~\ref{fig:exp-comparison-shape-2} (D).

As demonstrated in Figure~\ref{fig:exp-comparison-shape-2} (A) and (B), the resolution of reconstructed internal surfaces, produced by the isosurface extraction algorithm, depends on the resolution $r$ of $\mathcal{S}^{\mathrm{gssdf}}$. 
To optimally synchronize the resolution between the internal and external meshes, it is crucial to select and train a detailed enough resolution $r$ for the generative model $\mathcal{G}$ and $\mathcal{S}^{\mathrm{gssdf}}$, based on the resolution of the target input mesh. 
Following this, we can generate fractured shapes with similar resolutions for the internal and external surfaces by employing a mesh simplification method to adjust the mesh resolution.
Note that the resolution of $r = 128$, with a time cost of 8.8 seconds, already provides sufficiently detailed internal fractured surfaces comparable to the external mesh for the squirrel's target shape. Additionally, increasing the resolution to $r = 256$ will extend the cost to 36.3 seconds.

\begin{table}[tb]
  \centering  
  \scalebox{0.63} {
  \begin{tabular}{rrrrrrr@{}}
    \toprule
    & Run-time & Training Time & Shape Type & Retain & Re-fracture \\
    &  &  & Orig. Mesh &  \\\midrule
     BEM \cite{hahn2016fast} & 15-40 mins. & 3.8-10.5 hours & CP-based & No  & Yes & \\\hline 
     Our method  & 1.08-8.72 sec. & - & CP-like & Yes& No   & \\\hline
    Sell\'{a}n et al. & \multirow{2}{*}{1-5 ms.} & \multirow{2}{*}{20-180 mins.} & \multirow{2}{*}{Cutting-based} & \multirow{2}{*}{Yes} & \multirow{2}{*}{Yes} & \\
    \cite{breakinggood2023} & & & & \\\hline
    Schvartzman & \multirow{2}{*}{31-339 ms.} & \multirow{2}{*}{-} & \multirow{2}{*}{Voronoi} & \multirow{2}{*}{Yes} & \multirow{2}{*}{No}   & \\
     et al. \cite{schvartzman2014fracture} & & & & \\\hline
     Huang et al. & \multirow{2}{*}{10-30 mins.} & \multirow{2}{*}{43-288 mins}& \multirow{2}{*}{CP-like} & \multirow{2}{*}{No} & \multirow{2}{*}{Yes}   & \\
     \cite{kanailab2018} & & & & \\
    \bottomrule
  \end{tabular}
  }
  \caption{Qualitative comparison among BEM brittle fracture simulation, our methods, Sell\'{a}n et al.'s, and Schvartzman et al.'s. Our method is an alternative option for generating crack propagation-based brittle fracture animation. Note that CP denotes crack propagation. Our method's runtime costs are demonstrated at resolutions of 128 and 256.}
  \vspace{-5mm}
  \label{tab:comparison}
\end{table}

For reference, we also provide the rendered surface result of Sell\'{a}n et al.'s (C) and brittle fracture simulation's (D) in Figure~\ref{fig:exp-comparison-shape-2}, which is generated by a similar collision condition.
In our experiments, we generated fragments that reproduce Sell\'{a}n et al.'s results using $13{\small,}000$ cage elements and a mode value of 10. 
Also, all results are rendered by flat and wireframe shading in Figure~\ref{fig:exp-comparison-shape-2}.

\begin{figure}[tb]
  \centering
  \includegraphics[width=.15\textwidth]{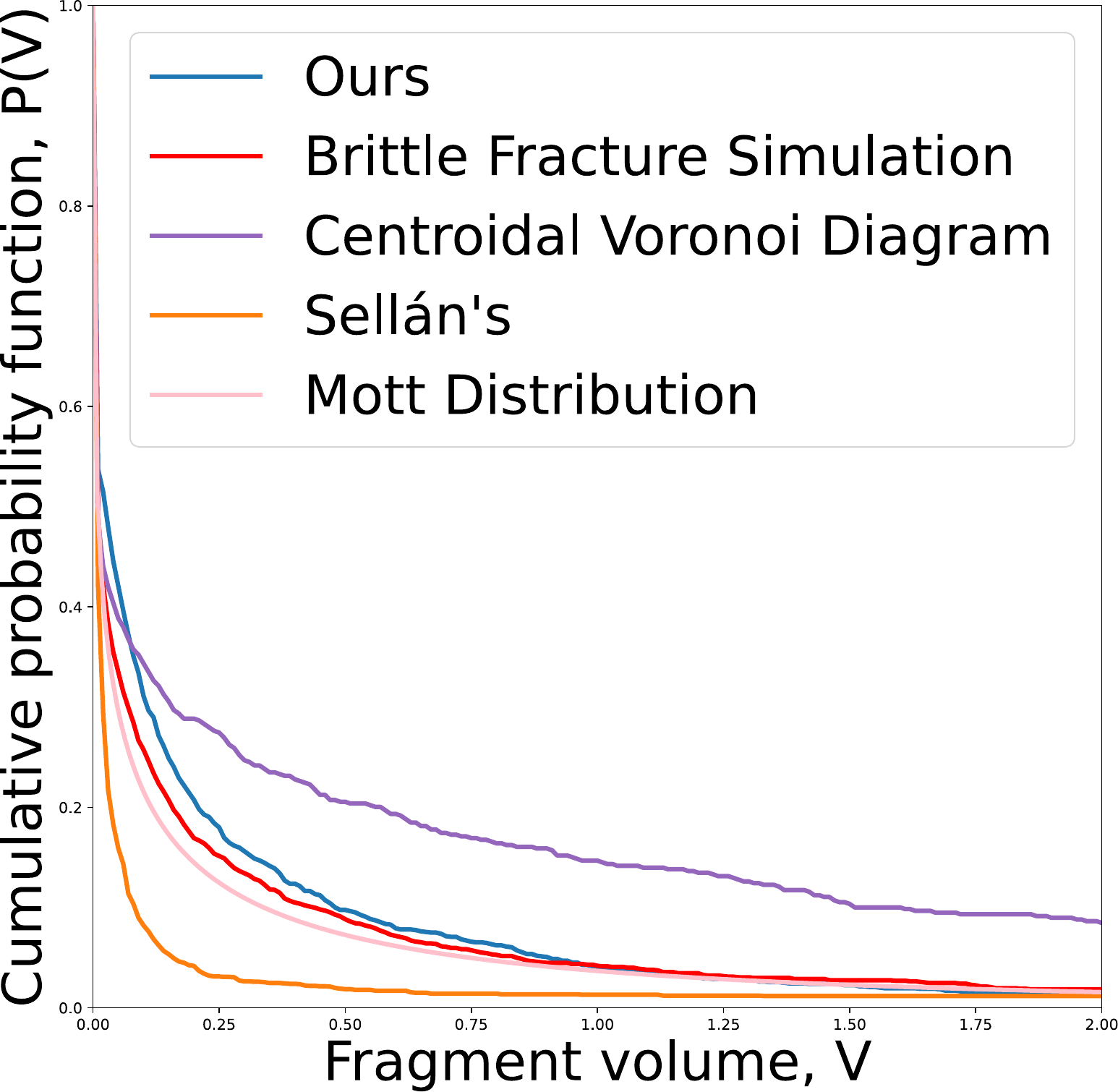}
\hfill
  \centering
  \includegraphics[width=.15\textwidth]{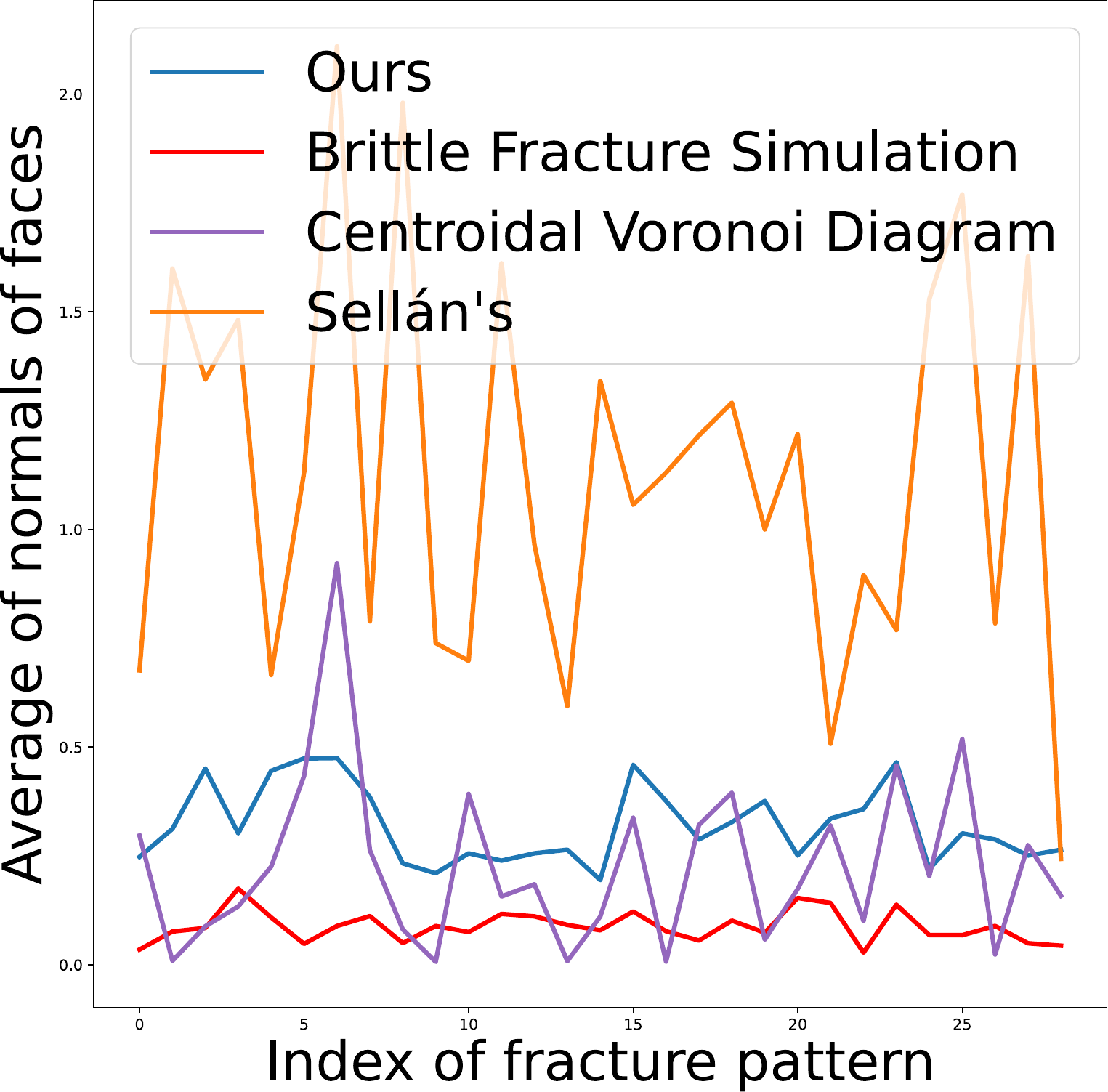}
\hfill
  \centering
  \includegraphics[width=.15\textwidth]{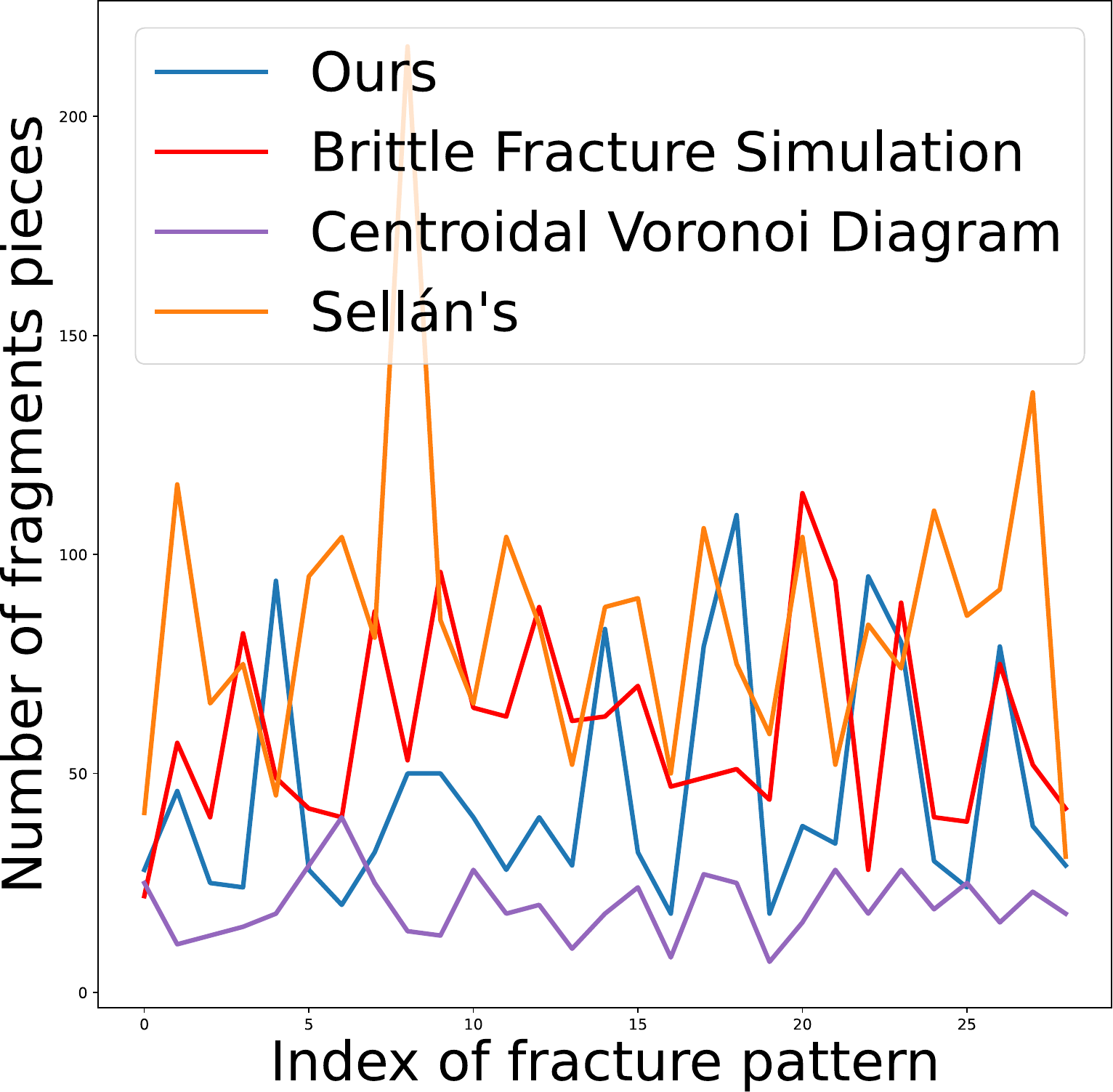}
\hfill
  \caption{
    Results of the quantitative comparison. Left: The distribution of fragment volumes; Middle: Average of fractured surfaces' normal; Right: Number of fragment pieces.
    }
    \label{fig:exp-quantity-compare}%
\end{figure}

Figure~\ref{fig:exp-quantity-compare} shows the quantitative comparison of our method, BEM fracture simulation, Sell\'{a}n et al.'s, and 3D CVD. Each method involved 30 collisions with the test cases. The method of the 3D CVD follows the implementing concept of Schvartzman et al. \cite{schvartzman2014fracture}. However, we did not implement the part of learning and provided handmade Voronoi control points to illustrate the fragments. 

We show the distribution of fragment volumes on the left of Figure~\ref{fig:exp-quantity-compare} and compare them to Mott’s formula. Mott’s formula is a widely accepted model for fragment size distribution, which is given by $P(V)=e^{-\sqrt[3]{\zeta V}}, \zeta = \frac{6}{\Bar{V}}$, where $\Bar{V}$ is the average volume of 30 times BEM fracture simulations (see also Equation (20) in \cite{elek2008fragment}). 
The distribution of fragment volumes is calculated as a cumulative probability function. The volume of the original mesh of the Squirrel is normalized to 10. The number of fragments and average surface normal are summarized in the middle and right of Figure~\ref{fig:exp-quantity-compare}.

The distribution of fragment volumes shows that our method can provide similar fragment volumes, surface normals, and the number of fragments compared with the other methods, While the result of Sell\'{a}n et al. tends to produce more fragments and coarse surfaces and the result of 3D CVD tends to produce fewer and bigger fragments.

Our method focuses on capturing the characteristics of instantaneous brittle fracture processes and the relationship between the collision situation and the crack propagation of shape fracture. Therefore, we do not consider the re-fracture of already broken shapes.
Both our method and Sell\'{a}n et al.'s can retain the surfaces of the original mesh. Table~\ref{tab:comparison} summarizes the benefits and drawbacks of each method. Our method is faster than crack propagation-based brittle fracture simulation but cannot reach cutting-based or Voronoi-based real-time methods, making it an alternative option for generating crack-propagation-like brittle fracture animation.

With the design of vector quantisation, our method can predict fracture patterns for arbitrary unseen collisions with varying positions, directions, and magnitudes of impulse strength.
Additionally, the supplemental material includes further visual comparisons involving different positions, directions, and magnitudes of impulse strength, as well as an ablation study involving different legacy networks.

\begin{table*}[tb]
  \centering
  \scalebox{0.67} {
  \begin{tabular}{@{}r|r|rrrrrrrr|rr@{}}
    \toprule
    & BEM \cite{hahn2016fast} & \multicolumn{8}{c|}{Our method}  & \multicolumn{2}{c}{Sell\'{a}n et al.\cite{breakinggood2023}} \\
    & Sim. & Data Gen. & Train. & Pred. & MorphSeg.  & MeshBool.  & \textbf{Recon.} & Others. & \textbf{Run-time} & Mode Gen. & Impact Proj.\\
    \midrule
    Pot         & \textbf{16.9 mins} & 2 days & 17.8 hours & 0.30s (0.35s) & 2.25s (17.17s) & 3.71s (11.31s) & \textbf{6.26s(28.83s)} & 33.74s (41.17s) & \textbf{40.0s (70.0s)} & 38.8 mins & \textbf{0.005s} \\
    Bunny & \textbf{28.8 mins} & 4 days & 30.4 hours  & 0.27s (0.25s) & 1.55s (15.90s) & 3.88s (8.03s) & \textbf{5.7s (24.18s)} & 43.3s (50.82s) & \textbf{49.0s (75.0s)} & 34.4 mins & \textbf{0.002s} \\
    Base & \textbf{13.5 mins} & 2 days  &  23.0 hours   & 0.30s (0.41s) & 1.84s (15.52s) & 1.92s (8.09s) & \textbf{4.06s (24.02s)} & 33.64s (51.08s) & \textbf{37.7s(75.1s)} &  17.0 mins & \textbf{0.001s} \\
    Squirrel    & \textbf{15.5 mins} & 3 days & 23.9 hours & 0.22s (0.23s) & 2.49s (18.22s) & 6.01s (18.14s) & \textbf{8.72s (36.59s)} & 64.28s (87.21s) & \textbf{73.0s(123.8s)} & 16.3 mins &\textbf{0.004s} \\
    \midrule
    Mean        &  \textbf{18.7 mins}  & 3 days & 23.8 hours & 0.27s (0.31s) & 2.03s (16.7s) &	3.88s (11.40s) & \textbf{6.19s (28.41s)} & 43.74s (57.57s) & \textbf{49.93s (85.98s)} & 26.6 mins & \textbf{0.003s} \\
    \bottomrule
    \end{tabular}}
  \caption{
  Comparison of computation times in Figure~\ref{fig:exp-compare-1}, where the values in brackets mean the time cost of higher resolution $r=256$ of $\mathcal{S}^{\mathrm{gssdf}}$ while the values without brackets are related to the resolution $r=128$:
  Sim.: Time required to generate a single simulation.
  Data Gen.: Time required to perform 200 simulations on the one machine described in Section \ref{sec:exp-dataset}.
  Train.: Network training time.
  Pred.: Time required to predict a single fracture pattern during the inference phase of the network.
  MorphSeg.: Time required to morphological segmentation with MorphoLibJ\cite{morpholibj2016} during run-time.
  MeshBool.: Time required to Boolean set operations in Section~\ref{sec:caged-sdf} during run-time.
  Recon.: Time required to reconstruct a three-dimensional fracture shape for a single fracture pattern during run-time.
  Others: Time required to meshes saving, loading, software loading, and rigid-body simulation at run-time.
  Run-time: Time required to produce a fracture animation with rigid-body simulation during run-time.
  Mode Gen.: Time required to pre-compute force and generate the fracture modes.
  Impact Proj.: Time is required to project the fractured shapes based on force information from the generated parent fracture pattern during a collision.
  }
  \vspace{-5mm}
  \label{tab:accuracy}
\end{table*}

\subsection{Training Generative Model with Different Resolutions in the Same Network}
\label{sec:multiple-resolutions}

To provide results across multiple resolutions, we train our models concurrently at different resolutions ($r = 32, 64, 128, 256$) as an optional training method. 
As illustrated in the ``Code-to-Voxel Autodecoder'' network shown in Figure \ref{fig:learning-process}, our network generates GS-SDFs at these varied resolutions.
Usually, we specify one output as our training target. By backpropagating through multiple outputs in a single update cycle, we ensure that outputs at different resolutions share the same discrete latent code $\mathbi{Z}_d$, enabling us to generate the same fracture patterns at various resolutions.
Compared to a single-resolution generative model, the multi-resolution model requires more time to generate additional resolutions of the GS-SDF. However, both single and multi-resolution models have similar network sizes, typically ranging from 185MB to 190MB, with storage increasing by about 1\% on average. The inference time for the multi-resolution model is similar to that of the single-resolution model, as shown in Table \ref{tab:accuracy}.
Once trained, the multi-resolution generative model can process a collision condition. The purpose and benefit of designing such a multi-resolution model is to offer users a choice in selecting the mesh resolution for fractured surfaces during the run-time process.
In the supplemental material, we provide the results of the three different resolutions with detailed surfaces generated by a multi-resolution generative model. 

\subsection{Destruction Animation Generation Time during Run-time}

For the results shown in Figure~\ref{fig:exp-compare-1}, Sell\'{a}n et al. took an average of 26.6 mins to create 5-30 destruction patterns for one shape with 6000 cages and 0.003 seconds to adapt the destruction shape at run-time. Our method took an average of 3 days on one machine to perform 200 destruction simulations. We trained a generative model on 200 patterns for one shape and generated a destruction pattern for any collision in an average of 6.19 seconds for resolution $r = 128$ and 28.41 seconds for resolution $r = 256$ at run-time.

Our deep learning-based method generated visually close results to the brittle fracture simulation in an average reasonable calculation time of 6.19 seconds, compared to the average calculation time of 18.7 minutes for the crack propagation-based brittle fracture simulation.

\begin{figure}[tb]
  \centering
  \includegraphics[width=\linewidth]{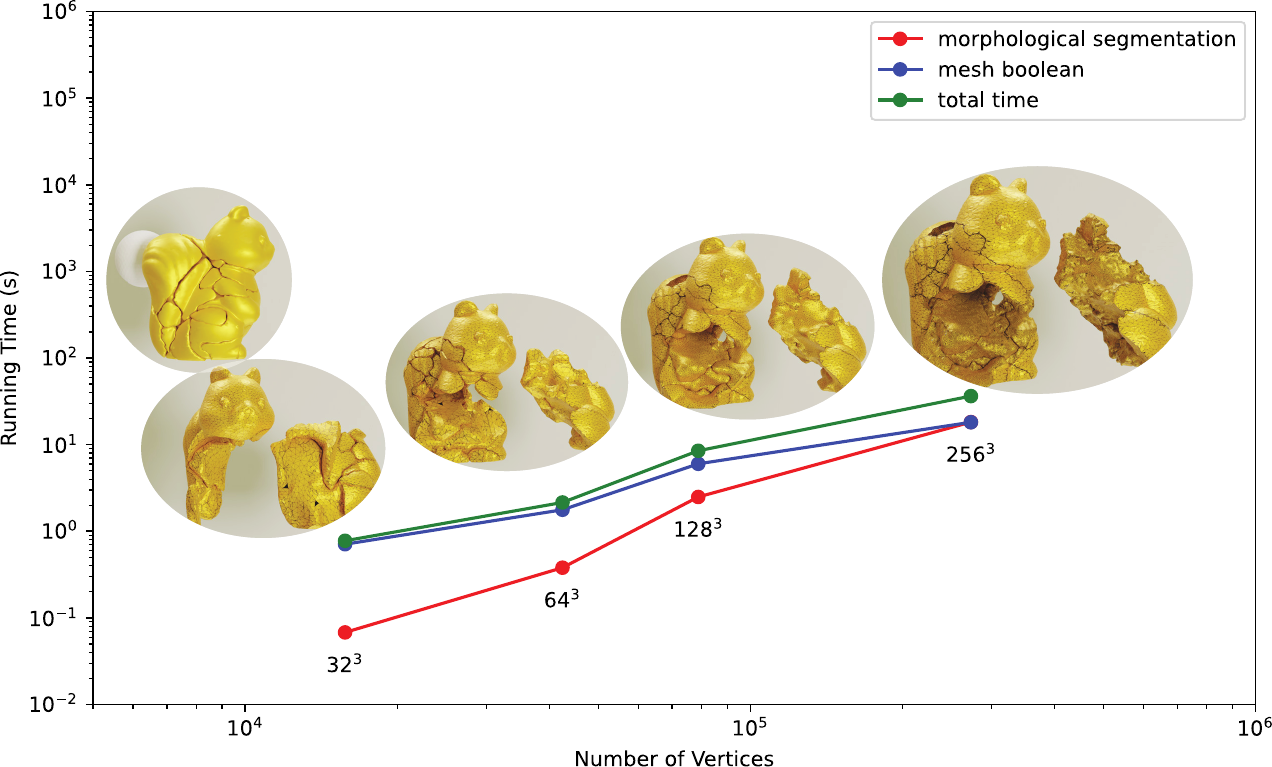}
  \vspace{-5mm}
  \caption{
    Comparison of computational time with different resolutions:
We used the same input as in the Squirrel case and generated both animations and surfaces at different resolutions with our multi-resolution model. Note that the computational time of the mesh boolean operation is mainly dependent on the number of vertices; the computational time of morphological segmentation is dependent on the voxel resolutions. Left to Right: voxel resolutions $r = 32, 64, 128, 256$.
    }
  \vspace{-5mm}
    \label{fig:exp-seg-bool-spped}%
\end{figure}

In addition, we compare the inference time with different resolutions for the squirrel shape at $r = 32,64,128,256$. We summarize the time of
morphological segmentation and Boolean set operation in Figure~\ref{fig:exp-seg-bool-spped}.
Figure~\ref{fig:exp-seg-bool-spped} shows that the time of mesh boolean operation is related to the number of vertices, and the time of morphological segmentation is related to the resolution of voxel grids.
As a result, the squirrel model with a resolution of $r = 32$ and $10^5$ vertices requires approximately 1 second of computational time.

\subsection{Limitations and Future Work}
\label{sec:limitation}

Our method faces challenges in implementing our technique for predicting various shapes and materials with general generative networks without re-training.

Even though our prediction can rapidly produce fracture patterns, it is hard to reach the real-time level.
As indicated by the reconstruction time costs for ``MorphSeg'' and ``MeshBool'' in Table~\ref{tab:accuracy}, our method currently struggles to generate fractures within half a second. However, our method can potentially achieve an interactive network-based fracture animation system in real-time with lower resolution in industry techniques.

Although our method shows the ability to handle multiple shapes in the single generative model, the quality of reproducing fractured shapes decreases as the number of diverse target shapes increases. 

Lastly, enhancing the resolution of voxel data poses difficulties due to the increased inference time of networks and morphological segmentation. This is further investigated when we try to augment network size and segmentation to refine prediction result details. 

In this paper, we train the generative model individually for each specific shape with a supervised small dataset. Looking ahead, addressing more general shapes with the current method in a single model proves impractical due to the extensive network size and data volume. Nevertheless, managing the data by categories (e.g., cat, plane, chair, etc.) with a general generative model might be a feasible approach.
Building a general generative model to predict fracture-like shapes for arbitrary shapes without supervised simulation data would be another valuable and challenging task. It requires a shape code prediction module or fine-tuning to predict fracture patterns tailored to unsupervised arbitrary shapes.

Nonetheless, due to the present challenges above, several enhancements are needed for our approach:
\begin{itemize}
    \item Develop a general generative model to predict brittle fractures across a specific category, rather than focusing on a particular shape.
    \item Enhance our deep learning fracture animation system to support various attributes, encompassing different materials.
    \item Explore and develop further deep learning methods capable of processing generative geometric segmentation as implicit function representations in a generative network.
    \item Develop a real-time SDF-based cage-cutting method by substituting the flooding algorithm.
\end{itemize}

%% file: 10_conclusion.tex
\section{Conclusion}

In this paper, we introduce the prior application of a deep learning-based fracture system and define the task of 3D destruction shape generation. We also develop a novel and stable SDF-based cage-cutting method that can be adapted in other works.

Our proposed method predicts brittle fracture shapes using a 3D generative network with discrete representation prediction tailored for rigid body simulations and brittle fracture. Experimental outcomes show that, compared to traditional methods at run-time, our deep learning approach can generate stable, more intricate, and lifelike destruction forms in a practical computation time frame.

%% file: 12_appendix.tex
\section{Brittle Fracture Physics: Prerequisites and Characteristic}

Fragments formed by brittle fractures represent the chaos observed in the natural world. 
Although fragment distribution has similar characteristics due to the cause-and-effect relationships driven by similar external forces, the measurement of exact fractured shapes differs. This variation arises from the discrete initiation of cracks in brittle materials.

The details of our data-generation framework are provided in Algorithm~\ref{alg-data-generation}, and the run-time framework is in Algorithm~\ref{alg-run-time} for implementing the Bullet Physics. The figures and equations relate to the main context.
The details of the data generation scene are summarised below:

\begin{itemize}
    \item The concept from Hahn and Wojtan [8] is adopted, treating all instantaneous crack propagation within a single frame of the rigid-body system as completed. 
    \item The primary object of fragments can fracture further. The dataset's final fractured result is captured after multiple-step fracturing within one second (60 frames).
    \item To simplify and abstract the fracturing process, a collision scenario between a breakable target and an unbreakable sphere is used.
    \item The most significant impulse values on the surface are used as the prediction input when a collision has multiple contact points.
    \item Gravity is set to zero in the dataset creation scene. The framework regards impulse as the main factor in fracturing.
    \item The variables impacting the fracture process of breakable objects are restricted to the position and direction of the collision and the magnitude of the impulse on its surface. 
\end{itemize}

\begin{algorithm}[tb]
\caption{Impulse Based Rigid Body Fracture Simulation Loop}
\label{alg-data-generation}
\begin{algorithmic}[1]
\While {true}
    \State solve rigid body dynamics [fig.2a]
    \State get impulses $J$ of breakable rigid bodies [equ.1]
    \If {$\sigma_1 > \sigma_0$} [equ.2]
        \If {$I_{\mathrm{max}} < I_1$}
            \State $I_{\mathrm{max}} = I_1$ 
        \EndIf
        \While {true}
        \State get traction field by impulses $J$ $\rightarrow$ dataset [fig.2d]
        \State solve BEM fracture simulation [7]
        \State register breakable fragments into the rigid bodies list
        \State update mass and velocity for new rigid bodies
        \If {finish additional fractures}
            \State break
        \EndIf
        \State solve rigid body dynamics
        \State get impulses $J$ during self-collision
        \EndWhile
        \State capture fragments $\mathcal{S}^{\prime}$ $\rightarrow$ dataset 
 [fig.2d]
    \EndIf
    \State continue rigid body dynamics [fig.2f]
    \State record $I_{\mathrm{max}} \rightarrow $ Dataset
\EndWhile
\end{algorithmic}
\end{algorithm}

\begin{algorithm}[tb]
\caption{Deep Learning Run-time Fracture Animation Loop}
\label{alg-run-time}
\begin{algorithmic}[1]
\While {true}
    \State solve rigid body dynamics [fig.2a]
    \State get impulses $J$ of breakable rigid bodies [equ.1]
    \If {$\sigma_1 > \sigma_0$} [equ.2]
        \State record breakable target's mass and velocity [sec.5.1]
        \State predict fracture pattern $\mathcal{S}^{\mathrm{gssdf}}$ from impulses $J$ [equ.3]
        \State perform caged-sdf segmentation [sec.5.2]
        \State register unbreakable fragments into the rigid bodies list
        \State update mass and velocity for new rigid bodies [sec.5.3]
    \EndIf
    \State continue rigid body dynamics [fig.2e]
\EndWhile
\end{algorithmic}
\end{algorithm}

\section{Results of Dual-Collision Scene and Comparison of Impulses between Single and Dual Collision Scene}

\begin{figure*}[!b]
  \centering
  \mbox{} \hfill
  \includegraphics[width=\linewidth]{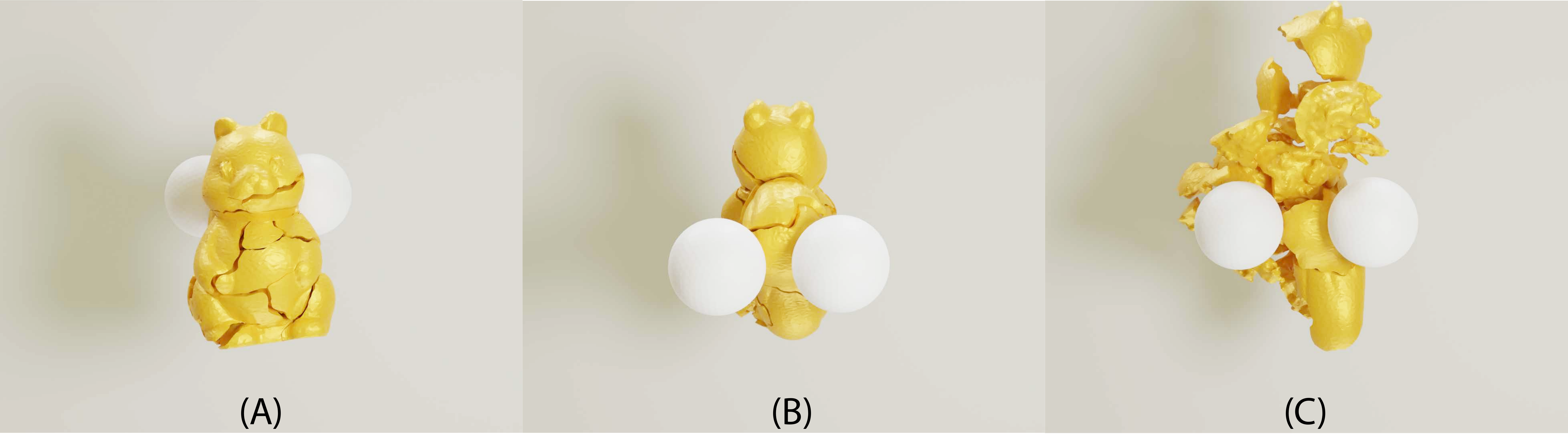}
  \hfill \mbox{}
  \caption{
    In the squirrel comparison, (A) and (B) share the same time while colliding, and (C) is 10 frames after the collision.
  }
  \label{fig:exp-squirrel-multi}
\end{figure*}

As shown in Figure~\ref{fig:exp-squirrel-multi}, our method can be extended to the multi-collision scene.
A collision scenario is designed, initiating two unbreakable spheres in random positions with random velocities and a breakable target in the shape of a Squirrel.
500 collision training data points are generated, and a model tailored to the Squirrel is trained.

Figure~\ref{fig:exp-squirrel-impact} compares a single collision scene with a dual collision scene.
Although the distribution of impulse strength differs, the largest impulse is larger than the second-largest.

\section{Evaluation with Scenes of Varying Magnitudes of Impulse Strength}

Scenes of balls shot in the same direction but with different magnitude of impulse strengths were examined for the Squirrel shown in Figure~\ref{fig:exp-squirrel-strength}. The Squirrel generative model created fracture patterns sensitive to impulse strength, producing reasonably complex and realistic fracture shapes appropriate for each collision.
The left column of Figure~\ref{fig:exp-squirrel-strength} shows the low-strength impulse collision across different frames. The impulse strength increases from left to right by raising the ball's initial velocity. 

As shown in Figure~\ref{fig:exp-squirrel-strength}, when comparing Ours A, Ours B, and Ours C, it can be observed that a larger increase in impulse strength results in smaller and more numerous predicted fragments.

\begin{figure*}[!b]
  \centering
  \mbox{} \hfill
  \includegraphics[width=\linewidth]{figs/exp-different-strength.pdf}
  \hfill \mbox{}
  \caption{
    In the Squirrel comparison, the varying collision strengths are presented from left to right: Case A (A)-(D), Case B (B)-(E), and Case C (C)-(F). Vertically, from top to bottom, we show the initial scene while colliding and 10-frames after the collision.
  }
  \label{fig:exp-squirrel-strength}
\end{figure*}

\section{Evaluation with Different Collision Scenes}

The generative models of Bunny, Base, and Squirrel were analysed to generate the examples in Figure 7 of the main text.

\begin{figure*}[!b]
  \centering
  \mbox{} \hfill
  \includegraphics[width=\linewidth]{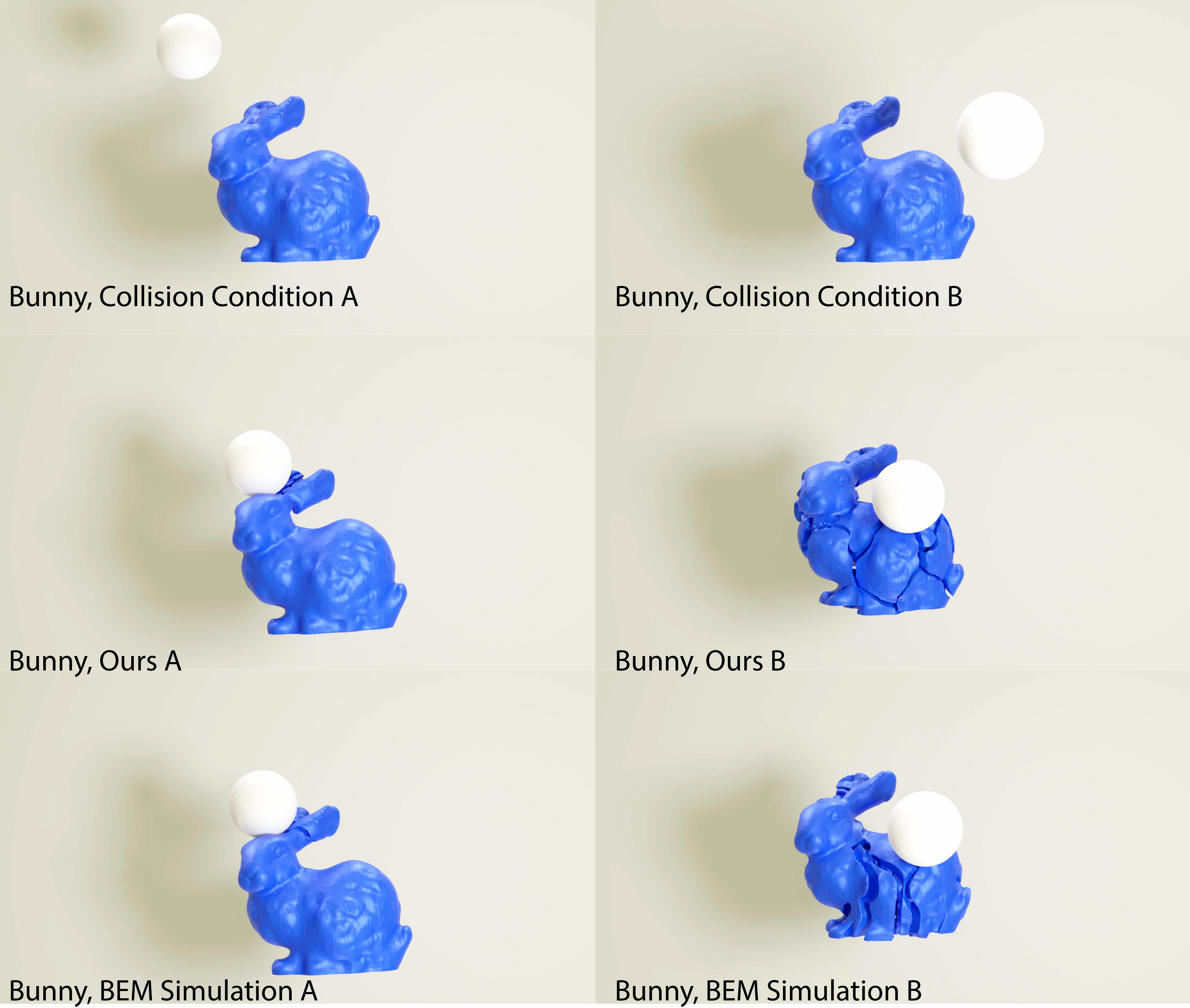}
  \hfill \mbox{}
  \caption{
    In the comparison for the Bunny, the varying collision directions are presented from left to right: Case A, Case B. Vertically, from top to bottom, we show the initial scene, followed by our method and then the BEM simulation. Each collision condition is not contained in the learning process.
  }
  \label{fig:exp-bunny-dp}
\end{figure*}

For the Bunny shown in Figure~\ref{fig:exp-bunny-dp}, we examined scenes of balls shot at different hit points. The Bunny generative model could reproduce fracture patterns focused on the part of shapes with the process described in Section 4.2.6 of the main text.

\begin{figure*}[!b]
  \centering
  \mbox{} \hfill
  \includegraphics[width=\linewidth]{figs/exp-base-different-directions.pdf}
  \hfill \mbox{}
  \caption{
    In the comparison for the Base, the varying collision directions are presented from left to right: Case A, Case B, and Case C. Vertically, from top to bottom, we show the initial frame of each scene, followed by our method and then the BEM simulation. All selected frames are taken post-collision. Each collision condition is not contained in the learning process.
  }
  \label{fig:exp-base-dd}
\end{figure*}

For the Base shown in Figure~\ref{fig:exp-base-dd}, we examined scenes of balls shot in different directions. Our Base generative model created fracture patterns sensitive to the impact position, producing reasonably complex and realistic fracture shapes appropriate for each distinct collision position. Nonetheless, the BEM simulation provides more realism by preserving large fragments on the side opposite the collision.

\begin{figure*}[!t]
  \centering
  \mbox{} \hfill
  \includegraphics[width=\linewidth]{figs/exp-squirrel-different-strength.pdf}
  \hfill \mbox{}
  \caption{
    For the Squirrel comparison, which examines impulses from varying hit parts but similar shooting directions, the sequence from left to right is: Case A, Case B, and Case C. Again, from top to bottom, we display the initial frame for each scene, our method, and the BEM simulation. Notably, Squirrel Collision B hits both the tail and head of the squirrel, while Squirrel Collision A is focused on the tail, and Squirrel Collision C is focused on the head. Each collision condition is not included in the learning process.
  }
  \label{fig:exp-squirrel-ds}
\end{figure*}

For the Squirrel in Figure~\ref{fig:exp-squirrel-ds}, we explored scenes with balls shot in a similar shooting direction but with varying hit points. In this figure, Squirrel Collision A and B represent the shot on the tail. Squirrel Collision B exhibits fractures in both areas, where the ball hits the squirrel's tail and head. Conversely, Squirrel Collision A shows fractures focused on the tail, where it was hit, while Squirrel Collision C displays fractures in the head. However, the random instance from our Squirrel did not replicate the same fractured shape as seen in the BEM simulation for the test case shown in Figure~\ref{fig:exp-squirrel-ds}.

\section{Evaluation with Near-shearing Impact Scene of Bar}

\begin{figure*}[!t]
  \centering
  \mbox{} \hfill
  \includegraphics[width=\linewidth]{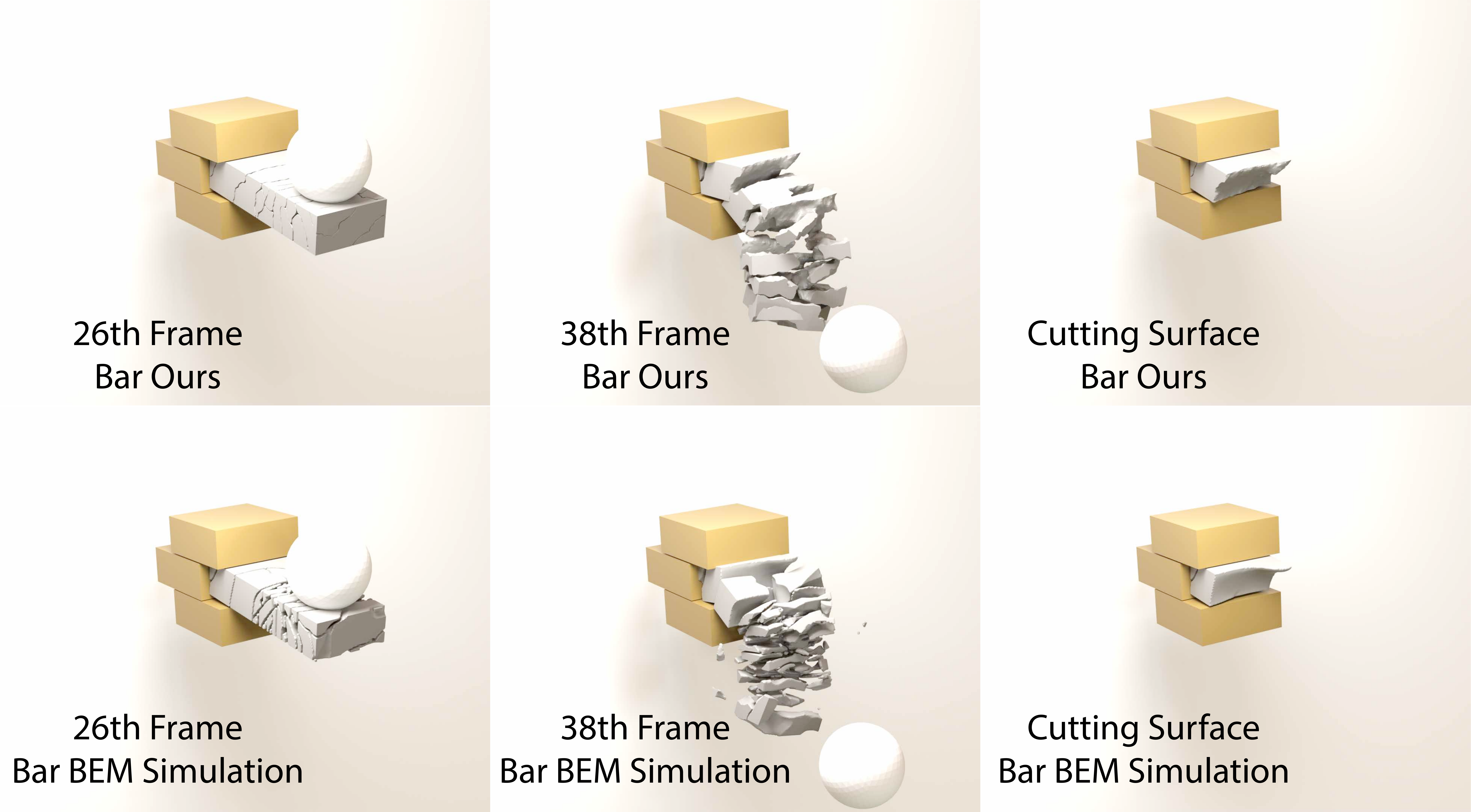}
  \hfill \mbox{}
  \caption{
    For the near-shearing evaluation, we examine the bar's shearing impulse. Left: the 26th frame while colliding; middle: the 38th frames after collision; right: the bar's cutting surfaces. Top row: the result of our method; Bottom row: the result of BEM fracture simulation.
  }
  \label{fig:exp-bar}
\end{figure*}

Figure~\ref{fig:exp-bar} shows the bar with a shearing impulse occurring during the collision. To provide the shearing cutting surfaces, we need to provide the near-shearing force and impulse scene during the scene of BEM crack-propagation simulations.
We proceed with the BEM simulation in the near-shearing scene to generate specific fractured shapes for the bar shape with 55 collisions. We train the generative model with the 50 collisions and test the new unknown test case of the collision scene with this model.

We need to note that it is not easy to generate a near-shearing force using the BEM simulation integrated with an impulse-based rigid-body system in [8] because the impulse-based information is hard to create scenarios like pulling, twisting, and cutting.
However, our method can provide the results if we give the training dataset scenarios.

\section{Justification of Voxel-based GS-SDF Representation and CNN-based Autodecoder}

\begin{figure}
  \centering
  \includegraphics[width=\columnwidth]{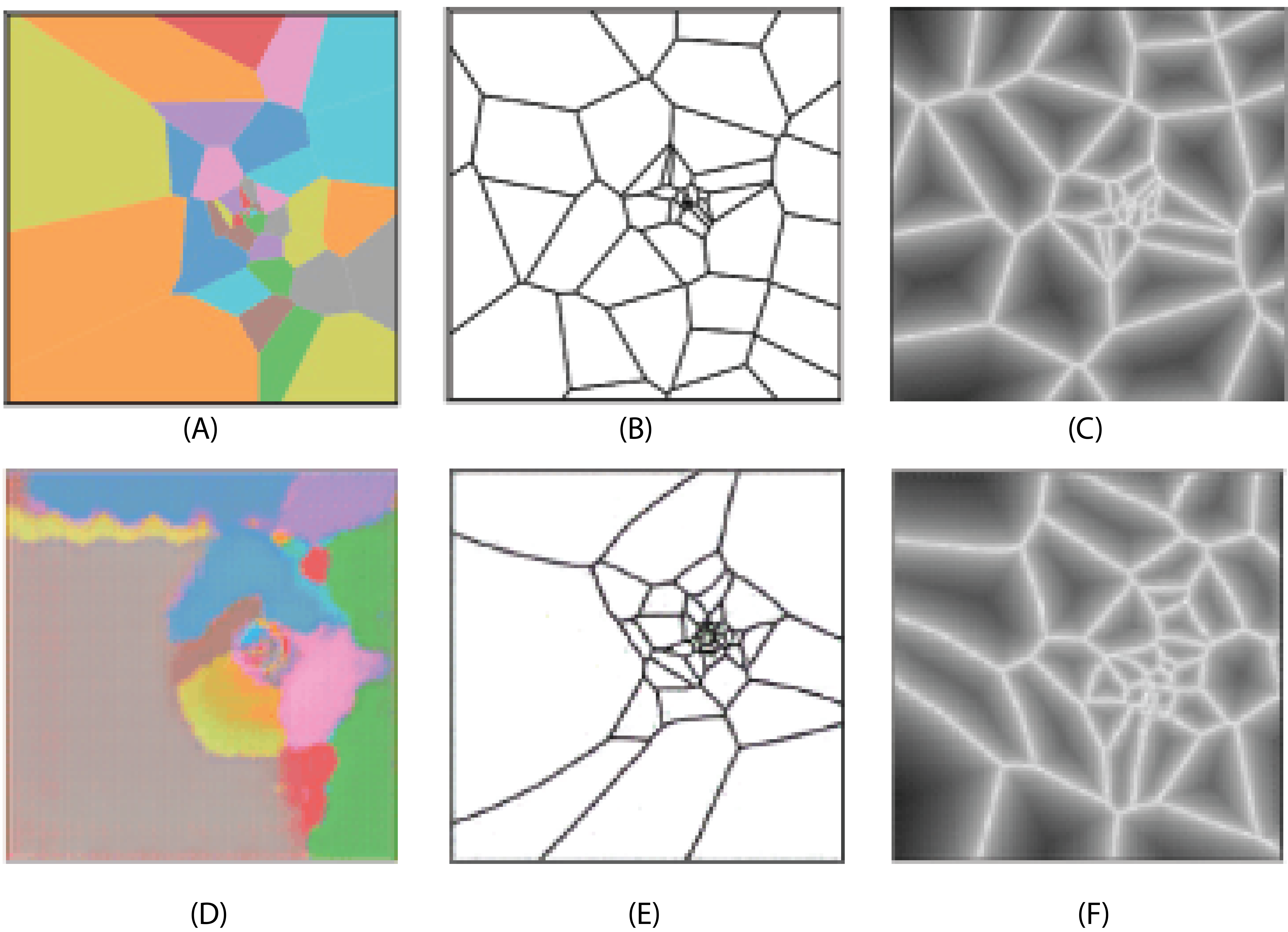}
  \caption{
  Comparison among the different representations of 2D GS-SDF. (A) Voronoi Diagram with Labels; (B) Voronoi Diagram with Lines; (C) Voronoi Diagram with 2D USDF; (D) Prediction Result of CNN Model Trained by Labeled data; (E) Prediction Result of CNN Model Trained by Line data; (F) Prediction Result of CNN Model Trained by 2D USDF data.
  }
  \label{fig:exp-2d-representation}
\end{figure}

\begin{figure}
  \centering
  \includegraphics[width=\columnwidth]{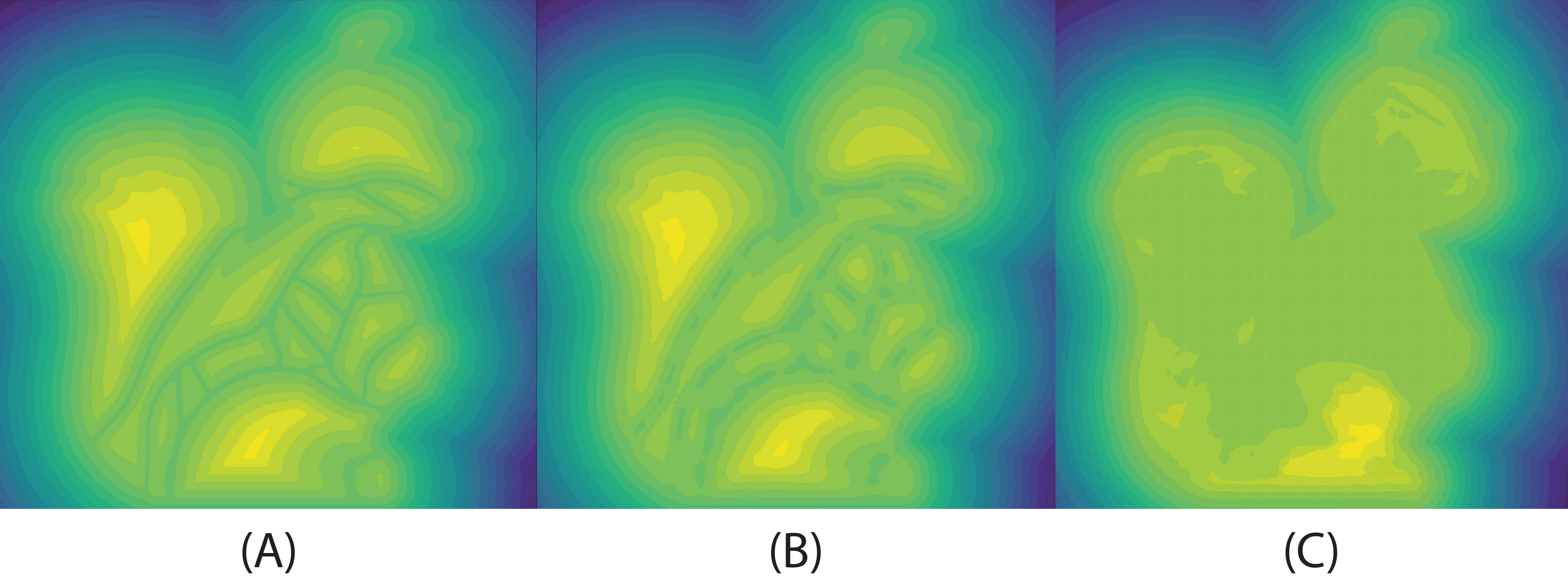}
  \caption{
  Comparison among the learning target of GS-SDF (A), voxel-based representation of $\mathcal{S}^{\mathrm{gssdf}}$ learned by the CNN-based autodecoder (B), and implicit function representation of $\mathcal{S}^{\mathrm{part}}$ learned by MLP-based autodecoder (C): 
  even though the MLP network can successfully learn the implicit function representation with the external surface of a watertight shape (pure-SDF), it fails to learn the representation of geometric segmentation with multiple adjacent fragments inside a watertight shape (GS-SDF) with autodecoder.
  }
  \label{fig:exp-network-compare}
\end{figure}

To justify using the SDF-based representation, we experiment with the 2D CNN model to be trained by label data, line data and USDF data in Figure~\ref{fig:exp-2d-representation}. This early stage shows that for the GS-SDF data, an SDF-based representation performs best in earning the stable border and distribution of area and size of the segments. We use the L2 loss function for the line and SDF-based data and dice loss in labelled data.
Unlike semantic segmentation and experiments in the medical image data, dice loss is not good in labelled data with variable number segments, and each region will be labelled into a random value but a semantic label.

To justify using the GS-SDF representation with a CNN-based autodecoder, we experiment using the MLP-based autodecoder designed by [26]. This experiment focuses on learning the implicit function representation of geometric segmentation. The segmentation involves multiple adjacent fragments within a watertight shape, defined by the mapping $(x, y, z) \mapsto s^{\mathrm{gssdf}}$. Here, $s^{\mathrm{gssdf}}$ represents a value sampled from $\mathcal{S}^{\mathrm{gssdf}}$.
The comparison results are shown in Figure \ref{fig:exp-network-compare}.

The geometric segmentation with multiple adjacent fragments differs from the pure SDF without a segmented region.
As shown in Figure \ref{fig:exp-network-compare},
the image of (A) represents the 40th depth of learning target of GS-SDF designed in Section 4.2.2 of the main text, which can be represented as a voxel-based representation.
Compared with the voxel-based representation of $\mathcal{S}^{\mathrm{gssdf}}$ learned by the CNN-based autodecoder as (B), we found that it is hard to reproduce the multiple geometric segmentation details during the inference phase inside an object while the external surface can be trained well with the MLP-based autodecoder. As mentioned in Section 4.2.3 of the main context, we share a similar training method with DeepSDF with a different network of MLP-based autodecoder as CNN-based autodecoder. 
The resolution in (B) is 128, and it visualises the 45th slice of voxel space. We sample a $(128, 128, 128)$ voxel space for the results from the MLP-based autodecoder, then extract and visualise the 45th slice as (C) in Figure~\ref{fig:exp-network-compare}.

Regarding the justification of the loss function, we need to note that, unlike the labelled mask or the one-hot mask representation used in 2D or 3D medical images, instance segmentation, and semantic segmentation, we will use cross-entropy loss or dice loss to represent a distinct area or region with a distinct value. Generative geometric segmentation needs to hold a variable number of segmented regions, and each region is homogeneous to the others, which means it cannot be attached with a fixed label for each area.
So, we used SDF-based representation with L2 Loss for GS-SDF representation. Also, we tried Eikonal loss and finally kept the L2 loss-only version in this paper.

\section{Training Loss}

\begin{figure}[tb]
  \centering
  \includegraphics[width=.23\textwidth]{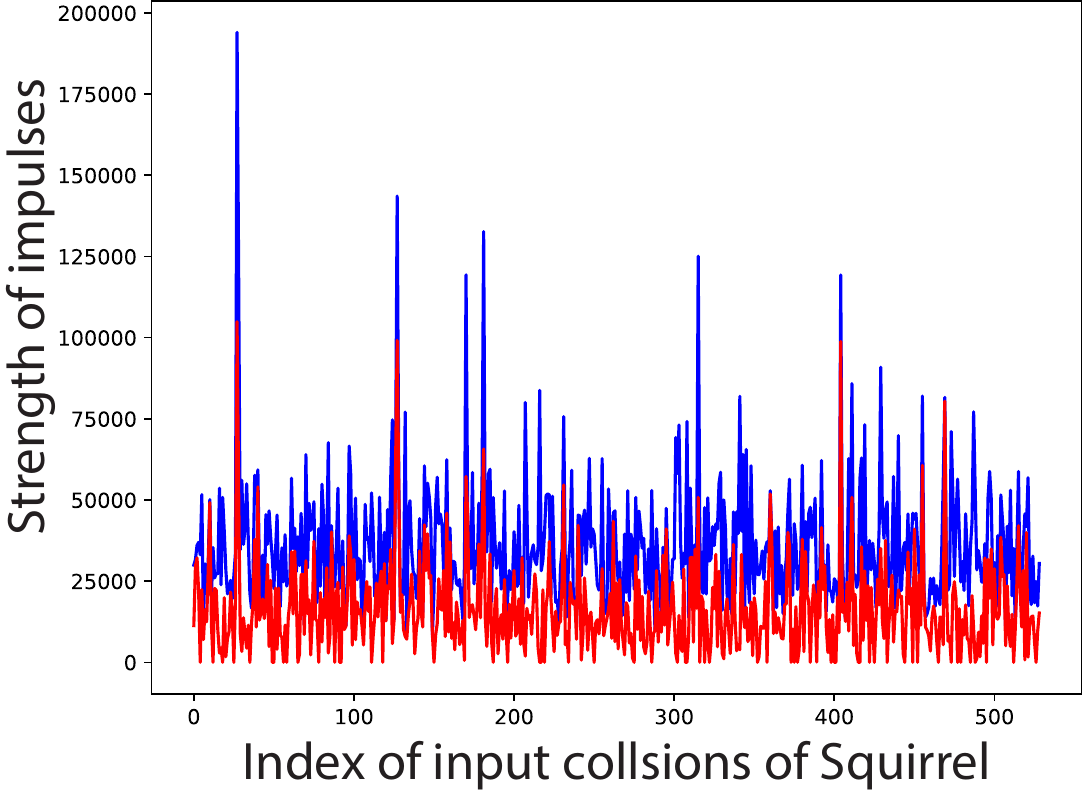}
\hfill
  \centering
  \includegraphics[width=.23\textwidth]{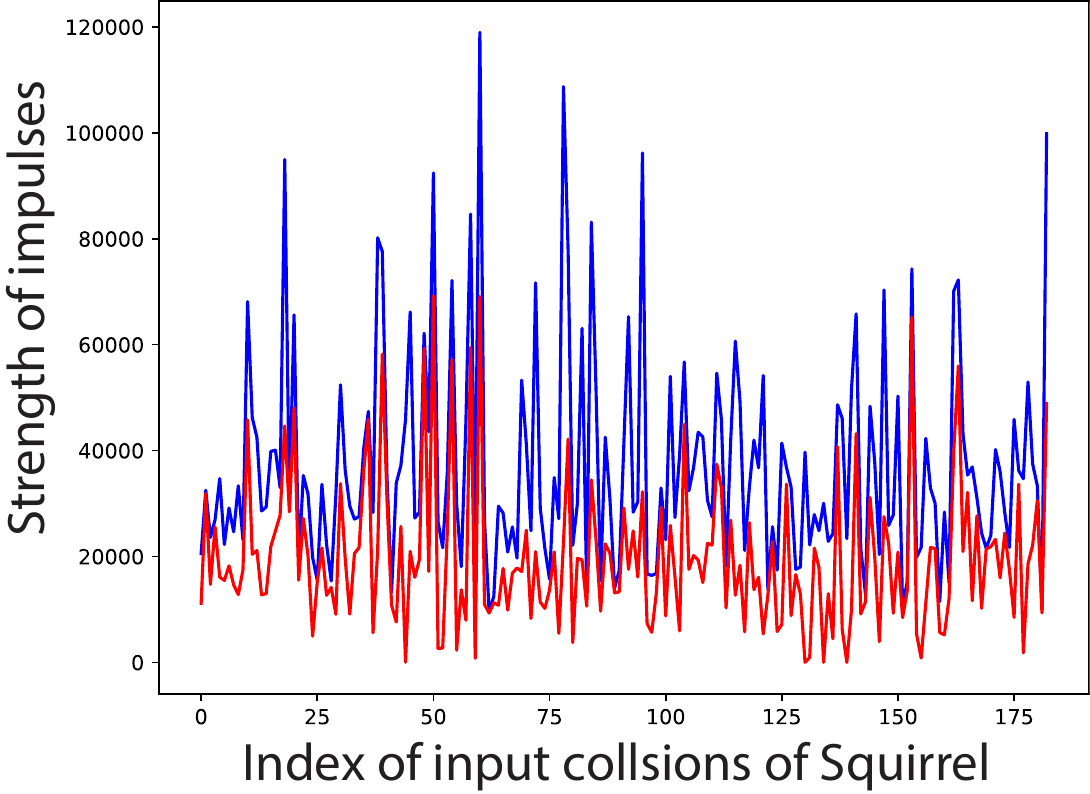}
\hfill
  \caption{
    Strength of impulses. Left: Scalar strength of impulses collected from the single collision scene; Right: Scalar strength of impulses collected from the dual collision scene; The blue line: the first-largest strength of impulse; The red line: the second-largest strength of impulse.
    }
    \label{fig:exp-squirrel-impact}
\end{figure}

\begin{figure}
  \centering
  \includegraphics[width=\columnwidth]{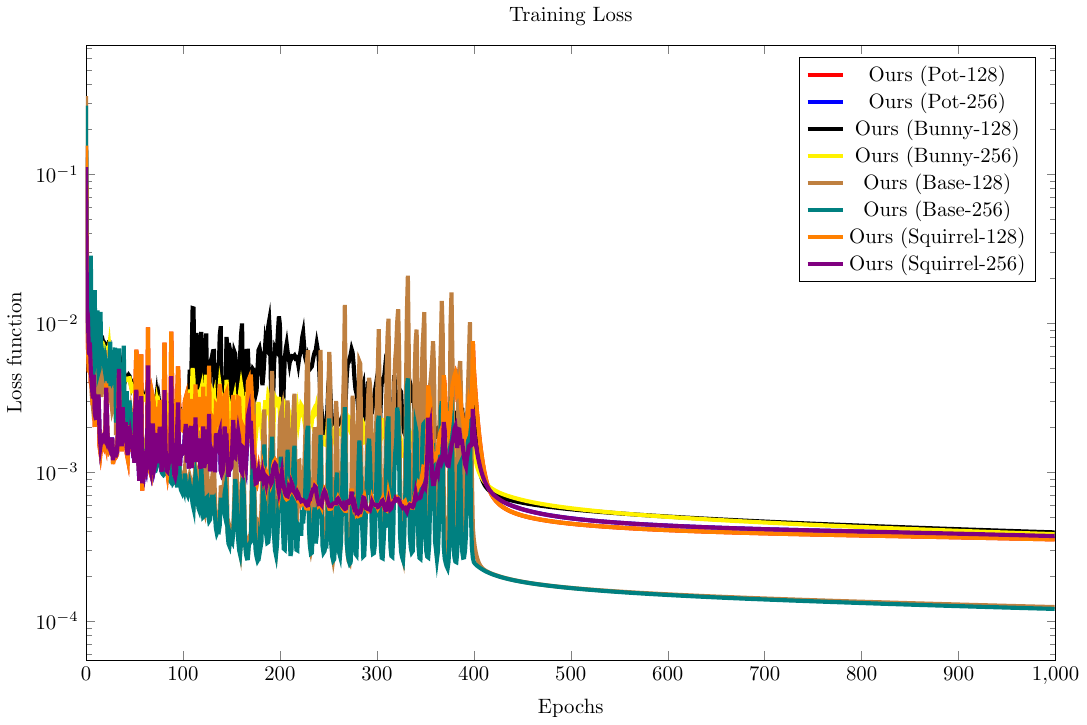}
  \caption{
  Training loss graph:
  We calculated and collected the $L_2$ loss while training the models shown in Figure 7 of the main text. The $L_2$ loss represents the objective in Section 4.2.5. It measures the model's ability to reproduce the fractured shapes' resolution. After the 400th epoch, we change the learning rate from 0.003 to 0.00005, which reduces the loss gradually after the 400th epoch.
  Ours (Pot-128, Squirrel-128, Bunny-128, Base-128): The proposed models with the resolution of $r=128$ for $\mathcal{S}^{\mathrm{gssdf}}$.
  Ours (Pot-256, Squirrel-256, Bunny-256, Base-256): The proposed models with the resolution of $r=256$ for $\mathcal{S}^{\mathrm{gssdf}}$.
  }
  \label{fig:exp-traning-loss}
\end{figure}

According to the Section 4.2.5 in the main text, we calculated and collected the
$L_2$ loss while training the models shown in Figure 7 of the main text. 
The results are shown in Figure \ref{fig:exp-traning-loss}.
Note that, as we observed, after the loss value is lower than 0.001, we can generate a stable fracture pattern with a discrete latent vector.
The minimum loss of each generative model is related to the size of the shape volume in voxel space, where the Base is the smallest shape among the target shapes.

\section{Results of Training Generative Model with Different Resolutions in the Same Network}

\begin{figure*}[tb]
  \centering
  \mbox{} \hfill
  \includegraphics[width=\linewidth]{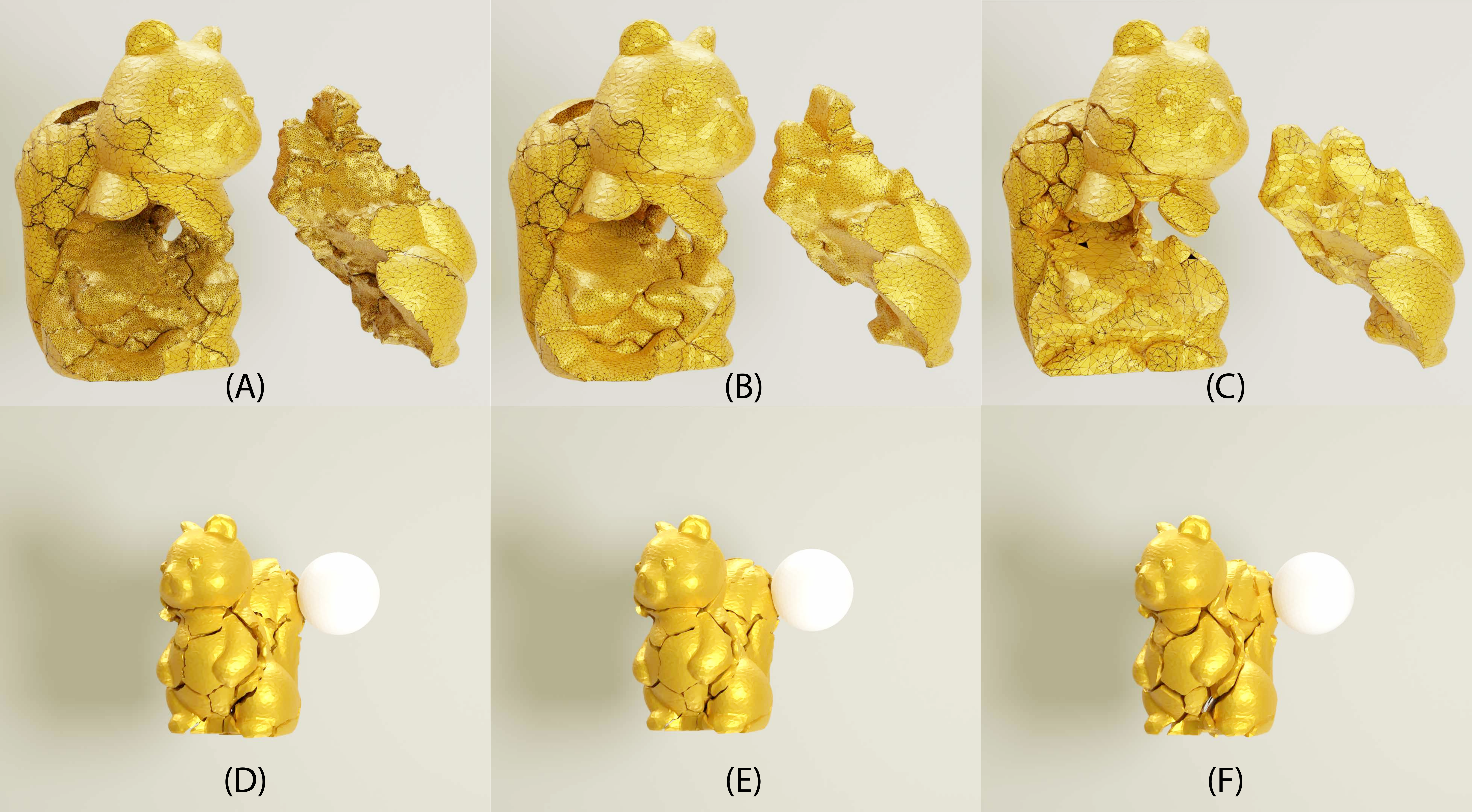}
  \hfill \mbox{}
  \caption{
    Results of the resolution comparison:
We used the same input as the Squirrel case and generated animations and surfaces with our multi-resolution model at different resolutions. Note that the mesh resolution corresponds to the voxel resolutions $r = 64, 128, 256$ used in training the generative model. (A) Squirrel animation at resolution $r = 256$; (B) Squirrel animation at resolution $r = 128$; (C) Squirrel animation at resolution $r = 64$; (D) Detailed surfaces in (A) at resolution $r = 256$; (E) Detailed surfaces in (B) at resolution $r = 128$; (F) Detailed surfaces in (C) at resolution $r = 64$.
    }
    \label{fig:exp-resolution}
\end{figure*}

To offer corresponding results across multiple resolution versions, we employ an optional training method that simultaneously trains the model at resolutions $r = 64, 128, 256$.
To demonstrate our generative model's capabilities, we compare animations and surfaces for the Squirrel example, using our multi-resolution model trained at these different resolutions, as shown in Figure~\ref{fig:exp-resolution}.

Note that the result of (F) in $r=64$ costs 2.2s for generating the fractured shapes in run-time and is acceptable to illustrate the fragment shapes in (C).
The result (E) in $r=128$ costs 6.1s for generating the fractured shapes in run-time and is detailed enough to illustrate the surface of fragments in (B).
The result (D) in $r=256$ costs 33.8s for generating the fractured shapes in run-time and illustrates the surface of fragments in (A).

\section{Ablation Study}

\begin{figure*}[tb]
  \centering
  \mbox{} \hfill
  \includegraphics[width=\linewidth]{figs/exp-ablation-study.pdf}
  \hfill \mbox{}
  \caption{
    Results of the ablation study:
    We chose the same input as the Squirrel case in Figure 7 of the main text and tested the case with several networks individually. Note that the test case's input is a ball collision with the Squirrel's tail. We expect the models to learn the high-density fragments surrounding the tail of the Squirrel, just like the result of the BEM simulation.
    Ours (Random C, Random D, Random E): The proposed models with different random normal distribution codes processed by Vector Quantization in the inference phase.
    w/o VQ (Random A, Random B): The model predicts directly with different random normal distribution codes without processing Vector Quantization with embedding space $\mathbbm{c}$.
    Pix2pix 3D: The 3D extended version of Pix2pix.
    w/o Segment Code: The model trained without normal distribution segment code.
    }
    \label{fig:exp-ablation-study}
\end{figure*}

We conducted an ablation study in Figure~\ref{fig:exp-ablation-study} to assess the efficiency of Vector Quantization, latent impulse representation of segment code, and the characteristic of random code.

We first compare our method with an enhanced 3D-Pix2pix based on Pix2Pix, where all 2D convolution layers are replaced with 3D convolution layers. Impulse information is encoded into voxels by appending the impulse scalar value to the voxel position. This method can produce fractured shapes similarly. However, as shown in Figure~\ref{fig:exp-ablation-study} (``Pix2pix 3D''), it struggles to represent fracture patterns with unknown new impulse input when collisions occur at the Squirrel's tail position.

Models without segment codes mean we do not use segment codes tailored to fracture patterns but use the normal distribution random codes in the latent impulse representation in the training phase, which means this model maps a noisy random representation to a fracture pattern. As shown in ``w/o Segment Code'', segment code provides a successful technique for achieving the learning task of connecting continuous latent code with discrete representation. Also, using a random code in run-time is inspired by the masking technique in generative networks [36]. Since we can not get the correct segment code, we can search for the correct latent impulse code by using a random code as the segment code to mask the code in a random normal distribution.

Pix2pix models (``Pix2pix 3D'') can only generate limited maps. Correlating these maps with the simulated fracture pattern and collision scenario is challenging.

Figure 4 of the main text shows that we initiate a normal distribution random code during the run-time process. This means that not only ``Random A'' but also ``Random B, C, D, and E'' in Figure~\ref{fig:exp-ablation-study} are produced. Nevertheless, without the process of Vector Quantization shown in Section 4.2.6 of the main text, there is a possibility of generating failure of too few fragments, as shown in w/o VQ (Random A). 
While some random codes fall near a specific discrete latent code in the embedding space, resulting in high-quality outcomes like Random B with many fragments reflecting the ``Collision on the tail'', the overall quality remains unstable without Vector Quantization.

Note that our method can provide stable outcomes using the vector quantization process. Random C, D, and E efficiently reflect the ``Collision on tail'' with a large number of fragments in the given context.
All GS-SDF outcomes in Random C, D, and E are stable. Feature distinct segmented region boundaries, as shown in (B) Figure~\ref{fig:exp-network-compare}. In contrast, results without normal distribution code (``w/o Segment Code''), Pix2pix (``Pix2pix 3D'') and ``Random A'' are reconstructed from unstable GS-SDF map.

\section{Training Time Comparison and More Test Cases}

We have tested our methods on various shapes, including the alphabet A, chair, lion, and mug.
Including the results in Figure 8 of the main context, we can summarize the training time of all models. 

We found that the training time of the learning process primarily depends on the dataset size. Once the loss value reaches 0.0004, the model can generate stable fracture shapes. The results presented in this paper include the maximum training time for stable fracture shape prediction, which is approximately 20 minutes.

\begin{figure*}[tb]
  \centering
  \mbox{} \hfill
  \includegraphics[width=\linewidth]{figs/exp-poster.pdf}
  \hfill \mbox{}
  \caption{
    Results of the other shapes. Left to Right: Results of Alphabet A, Chair, Lion, and Mug shapes; Top to Bottom: Input Collision Condition, Result of BEM Simulation, and Results of Our Method.
    }
    \label{fig:exp-poster}
\end{figure*}

\begin{figure*}[tb]
  \centering
  \mbox{} \hfill
  \includegraphics[width=\linewidth]{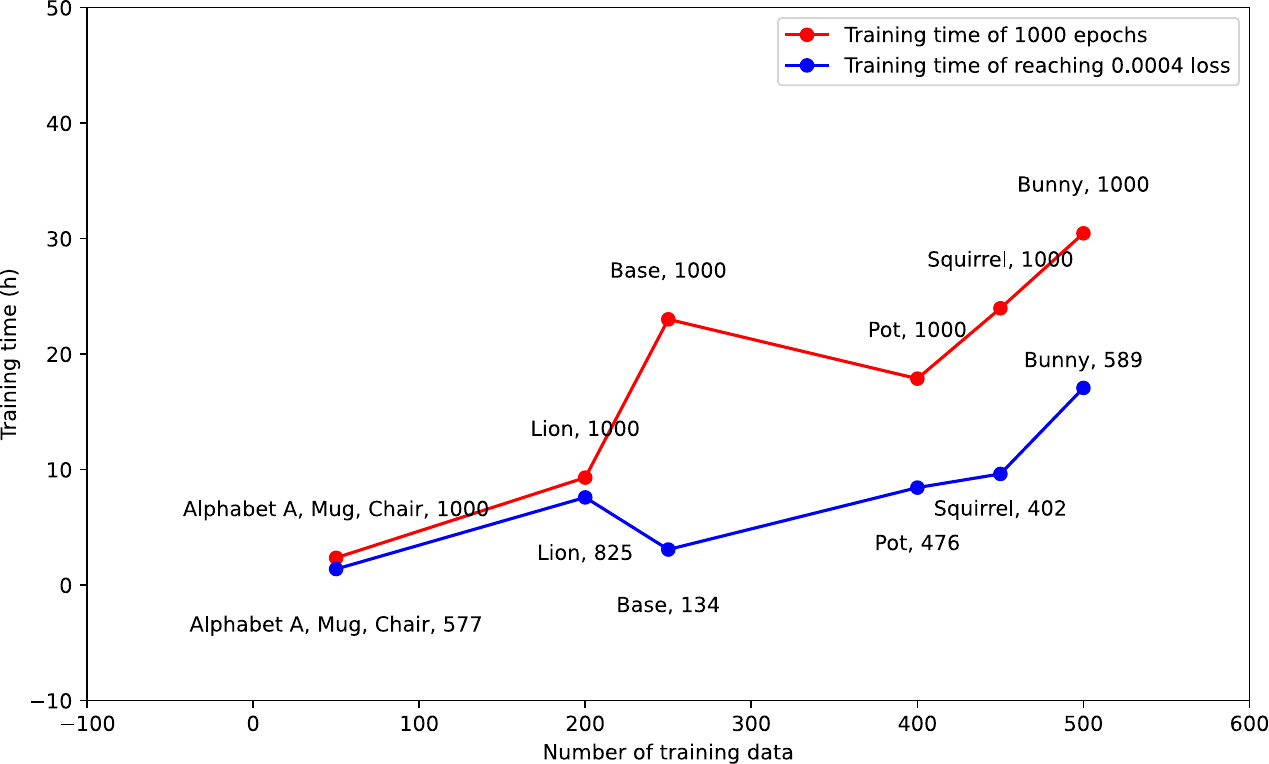}
  \hfill \mbox{}
  \caption{
    Results of the training time of Figure 8 in the main context and Figure~\ref{fig:exp-poster}. Besides the shape name, the number means the epochs in the training time.
    }
    \label{fig:exp-training-time}
\end{figure*}